\documentclass[twocolumn,aps,superscriptaddress,pre]{revtex4}

\usepackage{amsmath,amssymb,graphicx}
\usepackage{algorithmic}
\usepackage{enumerate}
\usepackage{times}
\usepackage{color}
\usepackage{soul}
\definecolor{yblue}{rgb}{0.06, 0.3, 0.57}
\usepackage[pdftex]{hyperref}
\hypersetup{colorlinks=true,linkcolor=blue,citecolor=blue,urlcolor=blue}

\begin{document}

\title{Evidence of many thermodynamic states of the three-dimensional Ising spin glass}

\author{Wenlong Wang}
\email{wenlongcmp@scu.edu.cn}
\affiliation{Department of Physics, Royal Institute of Technology, Stockholm, SE-106 91, Sweden}
\affiliation{College of Physics, Sichuan University, Chengdu 610065, China}

\author{Mats Wallin}
\affiliation{Department of Physics, Royal Institute of Technology, Stockholm, SE-106 91, Sweden}

\author{Jack Lidmar}
\affiliation{Department of Physics, Royal Institute of Technology, Stockholm, SE-106 91, Sweden}

\begin{abstract}
We present a large-scale simulation of the three-dimensional Ising spin glass with Gaussian disorder to low temperatures and large sizes using optimized population annealing Monte Carlo. Our primary focus is investigating the number of pure states regarding a controversial statistic, characterizing the fraction of centrally peaked disorder instances, of the overlap function order parameter. We observe that this statistic is subtly and sensitively influenced by the slight fluctuations of the integrated central weight of the disorder-averaged overlap function, making the asymptotic growth behaviour very difficult to identify. Modified statistics effectively reducing this correlation are studied and essentially monotonic growth trends are obtained. The effect of temperature is also studied, finding a larger growth rate at a higher temperature. Our state-of-the-art simulation and variance reduction data analysis suggest that the many pure state picture is most likely and coherent.
\end{abstract}

\maketitle

\textit{Introduction--} Spin glasses are fascinating disordered and frustrated magnets with a wide array of applications in diverse fields such as biology, computer science, and optimization problems \cite{Young:RMP,book}. The mean-field Sherrington-Kirkpatrick (SK) model \cite{SK} has an unusual low-temperature phase of many pure states described by the replica symmetry breaking (RSB) \cite{parisi:79,parisi:80,parisi:83}. Here, a pure state refers to a self-sustained thermodynamic state characterized by a time-averaged spin orientational pattern. Despite several decades of efforts, it is still an outstanding problem whether the more realistic Edwards-Anderson (EA) spin glass \cite{EA} in three dimensions has a single pair or many pairs of pure states. The RSB picture \cite{parisi:08,mezard:87} assumes that the mean-field theory is qualitatively correct for the EA model. On the other hand, the droplet picture \cite{mcmillan:84b,bray:86,fisher:86,fisher:87,fisher:88} based on the domain-wall renormalization group method predicts only a single pair of pure states much like a ferromagnet. The two pictures also differ on the geometrical aspect of excitations (fractals or space-filling) \cite{Wang:Fractal}, and the existence of a spin-glass phase in a weak external magnetic field \cite{almeida:78}. There are other pictures as well \cite{book}. The applicability of RSB is of broad interest and is related to, e.g., the Gardner transition in structural glasses \cite{charbonneau:14,Mike:GT}.


Despite much research aiming at discriminating which picture is suitable in three (and also four) dimensions, the problem has not been definitely settled. In this work we solely discuss the number of pure states, as a solid answer on one individual property can put stringent constraint on possible theories. There is mounting evidence that the disorder-averaged overlap function is nontrivial (corresponding to many pure states) for the sizes available, which have been steadily growing over time. One exception is the works focusing on the ground states at zero temperature \cite{Young:GS,Hatano:GS}. However, we conjecture that a single pair of ground states is a strong support for neither a two state nor many state picture. It seems likely what is going on in this case is that there are nonzero energy gaps between the lowest pure states. In this way, even $O(1)$ large-scale droplet excitations are forbidden at $T=0$. This is motivated by the observation that the disorder-averaged central weight [see Eq.~(\ref{I02})] decreases approximately linearly with decreasing temperature \cite{alvarez:10a}. 
Next, we turn to the finite temperatures, which is the focus of this work.


We find the computational controversies regarding the number of pure states are essentially from investigating new statistics, new boundary conditions, or a combination of the two. To our best knowledge, all numerical simulations find nontrivial disorder-averaged overlap functions under periodic boundary conditions at a typical low temperature. Therefore, to argue against many states, it is necessary that one or several of the conditions have to be altered. 
The free boundary condition was thought to potentially support a two state picture as $I(0.2)$ [see Eq.~(\ref{I02})] is found to rapidly decrease for small sizes and remarkably agrees with the $1/L^{\theta_{\rm{DW}}}$ scaling, where $\theta_{\rm{DW}}$ is the interface free energy exponent \cite{Katzgraber:FBC}. However, this was later found to be a finite-size effect from the surfaces \cite{Wang:FBC} and $I(0.2)$ of the free and periodic boundary conditions run together for larger sizes, supporting many pure states. This also suggests that the many pure states are genuinely due to droplet excitations rather than topologically protected domain walls.  By contrast, various statistics have been proposed other than $I(0.2)$, but most of these are not very successful; see, e.g., \cite{Yucesoy:Delta2} and the references therein for a collection of examples. One controversial but stimulating statistic is the fraction of centrally peaked instances \cite{Yucesoy:Delta,billoire:13,Yucesoy:Reply}, which we discuss in detail below.
For new statistics and boundary conditions, a work on sample stiffness in thermal boundary conditions argued against many states \cite{Wang:TBC}. This is also partially addressed \cite{Wang:EA4D}, and to fully resolve this problem a contrast experiment of the SK model should be conducted, which shall be discussed elsewhere.

In this work we focus on the controversial statistic $\Delta$ of the fraction of centrally peaked instances \cite{Yucesoy:Delta,billoire:13,Yucesoy:Reply}, and find again that there is no violation of many pure states. This is significant since $\Delta$ appears to do the best job among the new statistics supporting the two state picture \cite{Yucesoy:Delta2}. In \cite{Yucesoy:Delta}, it was found that $\Delta$ at $T=0.42$ grows with system size up to about $L=10$, then it levels off or stops growing appreciably. By contrast, $\Delta$ of the SK model at a similar $T/T_C$ grows prominently for the similar range of sizes. A criticism in \cite{billoire:13} suggested that comparing $T=0.4T_C$ for different models has no theoretical basis, and the difference is from the effective lower temperature of the EA model, i.e., a smaller central weight $I(0.2)$. Increasing the temperature of the EA model such that it has a similar $I(0.2)$ as the SK model, it was found that $\Delta$ also grows prominently in the EA model. However, the problem was not fully addressed in spite of the profound insight. An explanation of the irregular low temperature data is missing, and slightly different models were compared. The former group used Gaussian disorder and a low temperature \cite{Yucesoy:Delta}, while the latter group used $\pm J$ disorder and a relatively high temperature \cite{billoire:13}. 

The main purpose of this work is to resolve this problem systematically by carrying out a large-scale simulation of the three-dimensional EA model at the same parameters using the same model as the original work \cite{Yucesoy:Delta} but including larger sizes. In light of \cite{billoire:13}, data at a higher temperature are also collected for comparison. Using massively parallel population annealing Monte Carlo \cite{Hukushima:PA,Zhou:PA,Machta:PA,Wang:PA,Weigel:PA} and scalable MPI parallel computing, and taking further advantage of the recent optimizations of the algorithm \cite{Amey:PA,Amin:PA,Weigel:PA}, we have managed with considerable efforts to increase the largest size from $12^3$ spins \cite{Yucesoy:Delta} to $16^3$ spins. We refer to \cite{Amey:PA} for a discussion of how notoriously the spin glass computational complexity grows at low temperatures with the number of spins. Our larger range of sizes crucially enables us to identify a \textit{subtle} correlation between $\Delta(q_0,\kappa)$ and $I(q_0)$, showing that even small $I(q_0)$ fluctuations can significantly influence the behaviour of $\Delta$. Motivated by our observation, we define a slightly modified $\Delta$ compensating effectively for this correlation effect. The modified $\Delta$ essentially grows monotonically, providing a coherent many state picture. Our results also confirm that the different results of earlier works originate from the use of different effective temperatures or central weights \cite{billoire:13}. 



\textit{Model, method, and observables--} We study the three-dimensional Edwards-Anderson Ising spin glass \cite{EA} defined by the Hamiltonian $H = - \sum_{\langle ij \rangle} J_{ij} S_i S_j$, where $S_i=\pm 1$ are Ising spins and the sum is over nearest neighbours
on a simple cubic lattice under periodic boundary conditions. For a linear size $L$, there are $N=L^3$ spins. The random couplings $J_{ij}$ are chosen independently from the standard Gaussian distribution $n(0,1)$. 
A set of quenched couplings $\mathcal{J}=\{ J_{ij} \}$ defines a disorder realization sample or an instance. The model has a spin-glass phase transition at $T_C \approx 1$ \cite{katzgraber:06}. 

Population annealing cools gradually a population of $R$ random replicas starting from $\beta=0$ with alternating resampling and Metropolis sweeps until reaching the lowest temperature. In a resampling step, when the inverse temperature is increased from $\beta$ to $\beta'$, a replica $i$ is copied according to its energy $E_i$ with the expectation number $n_i=\exp[-(\beta'-\beta) E_i]/Q$. Here, $Q=(1/R)\sum_i \exp[-(\beta'-\beta) E_i]$ is a normalization factor to keep the population size approximately the same as $R$. In our simulation, the number of copies is randomly chosen as either the floor or the ceiling of $n_i$ with the proper probability to minimize fluctuations. After the resampling step, Monte Carlo sweeps are applied to each replica at the new temperature. Population annealing is used because it is both efficient and massively parallel \cite{Amin:PA,Wang:PA,Wang:GS}. Our equilibration criterion is based on the replica family entropy $S_f = -\sum_i f_i \log (f_i)$, where $f_i$ is the fraction of replicas descended from replica $i$ of the initial population. We require $S_f \geq \log(100)$ at the lowest temperature for each individual instance \cite{Wang:TBC,Wang:PA}. The preliminary simulation parameters are summarized in Table~\ref{table}, and unequilibrated instances were rerun with larger parameters until equilibration is reached. It should be noted that the hard instances may require substantially more work than a typical instance. For example, our typical top 5\% hard instances at $L=16$ require approximately $R\sim 10^7$ replicas, $N_T=501$ temperatures, and as large as $N_S=200$ sweeps on each replica per temperature; cf. the preliminary parameters in the table. Finally, our data readily pass the disorder average based equilibration test of \cite{katzgraber:01} at the lowest and a higher temperature ($T=2/3$) for all sizes.

\begin{table}
\caption{
Preliminary parameters of the population annealing simulation. $L$ is the linear system size, $R$ is the number of replicas, $T_0$ is the lowest temperature simulated, $N_T$ is the number of temperatures used in the annealing schedule, $N_S$ is the number of sweeps applied to each replica after resampling, and $M$ is the number of instances studied. Note that unequilibrated instances were rerun with (much) larger simulation parameters; see the text for details.
\label{table}
}
\begin{tabular*}{\columnwidth}{@{\extracolsep{\fill}} l c c c c r}
\hline
\hline
$L$  &$R$  &$T_0$ &$N_T$ &$N_S$ &$M$ \\
\hline
$4$  &$8\times10^4$ &$0.42$  &$101$ &$10$  &$5000$ \\
$6$  &$1.6\times10^5$ &$0.42$  &$101$ &$10$  &$5000$ \\
$8$  &$4\times10^5$ &$0.42$  &$201$ &$10$ &$5000$ \\
$10$ &$9.6\times10^5$ &$0.42$  &$301$ &$10$  &$5000$ \\
$12$ &$9.6\times10^5$ &$0.42$ &$301$ &$10$  &$5000$ \\
$14$ &$9.6\times10^5$ &$0.42$ &$401$ &$20$ &$3500$ \\
$16$ &$9.6\times10^5$ &$0.42$ &$401$ &$40$ &$3424$ \\
\hline
\hline
\end{tabular*}
\end{table} 

Our primary observable is the spin overlap distribution function $P_{\mathcal{J}}(q)$ where the spin overlap $q$ is defined as:
\begin{equation}
q=\dfrac{1}{N}\sum\limits_i S_{i}^{(1)} S_{i}^{(2)},
\label{q}
\end{equation}
where spin configurations ``(1)'' and ``(2)'' are chosen independently
from the Boltzmann distribution. We have collected the distributions at two typical low temperatures $T=2/3$ and $T=0.42$. Other regular observables like energy, free energy, and the replica family entropy are collected at all temperatures.

We further introduce two statistics of the individual overlap function: the central weight and central peakedness \cite{Yucesoy:Delta}:
\begin{eqnarray}
I_{\mathcal{J}}(q_0) = \int_{-q_0}^{q_0} P_{\mathcal{J}}(q) dq,
\label{I02}
\end{eqnarray}
\begin{eqnarray}
\Delta_{\mathcal{J}}(q_0,\kappa) = \theta\left( \max_{|q| \leq q_0}[(P_{\mathcal{J}}(q)+P_{\mathcal{J}}(-q))/2] - \kappa\right).
\label{D}
\end{eqnarray}
Here, $q_0$ characterizes the half length of a chosen interval around $q=0$, $\kappa$ is a chosen threshold to determine whether or not an instance is centrally peaked, and the Heaviside function $\theta(x)=1$ if $x \geq 0$ and $0$ otherwise. The statistic $\Delta_{\mathcal{J}}$ takes either $0$ or $1$. When the subscript is dropped, we refer to the disorder-averaged quantity. Hence, $\Delta$ refers to the fraction of centrally peaked instances. $\Delta$ should decrease to $0$ for a two state picture while it should increase to $1$ for a many state picture in the thermodynamic limit.



\begin{figure}[htb]
\begin{center}
\includegraphics[width=\columnwidth]{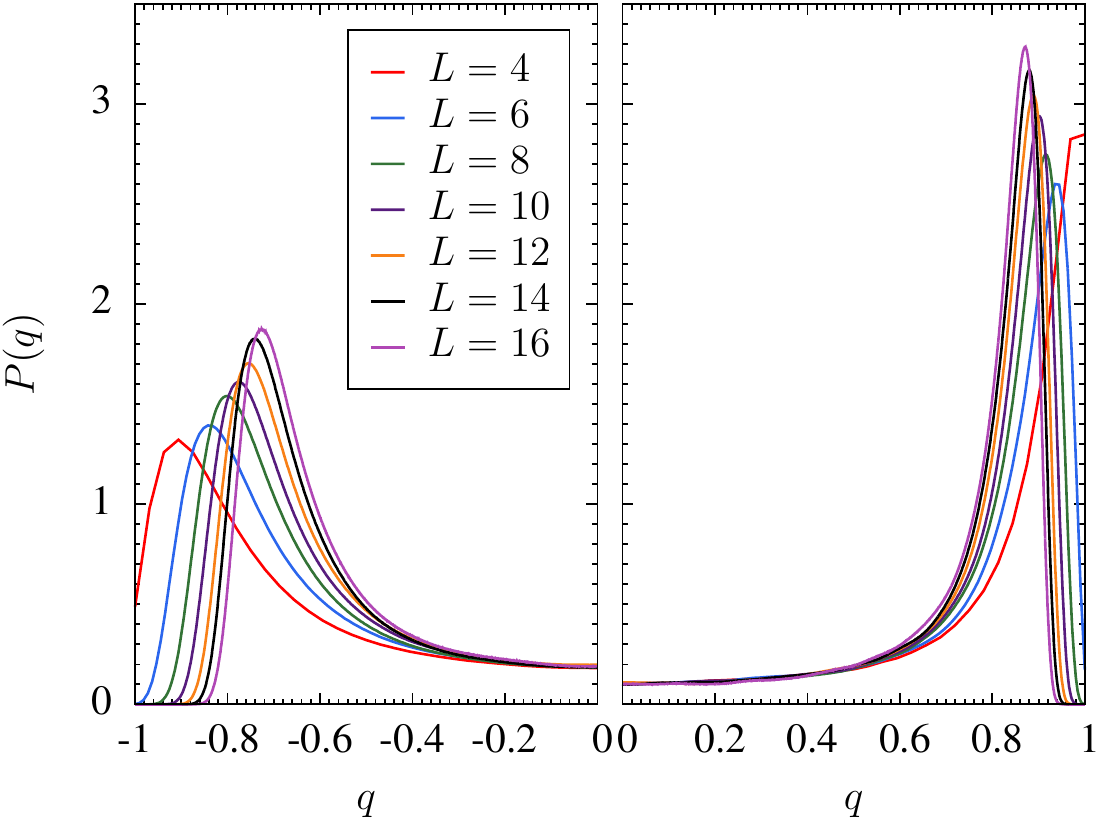}
\put (-211,170) {$(a)$} \\
\includegraphics[width=\columnwidth]{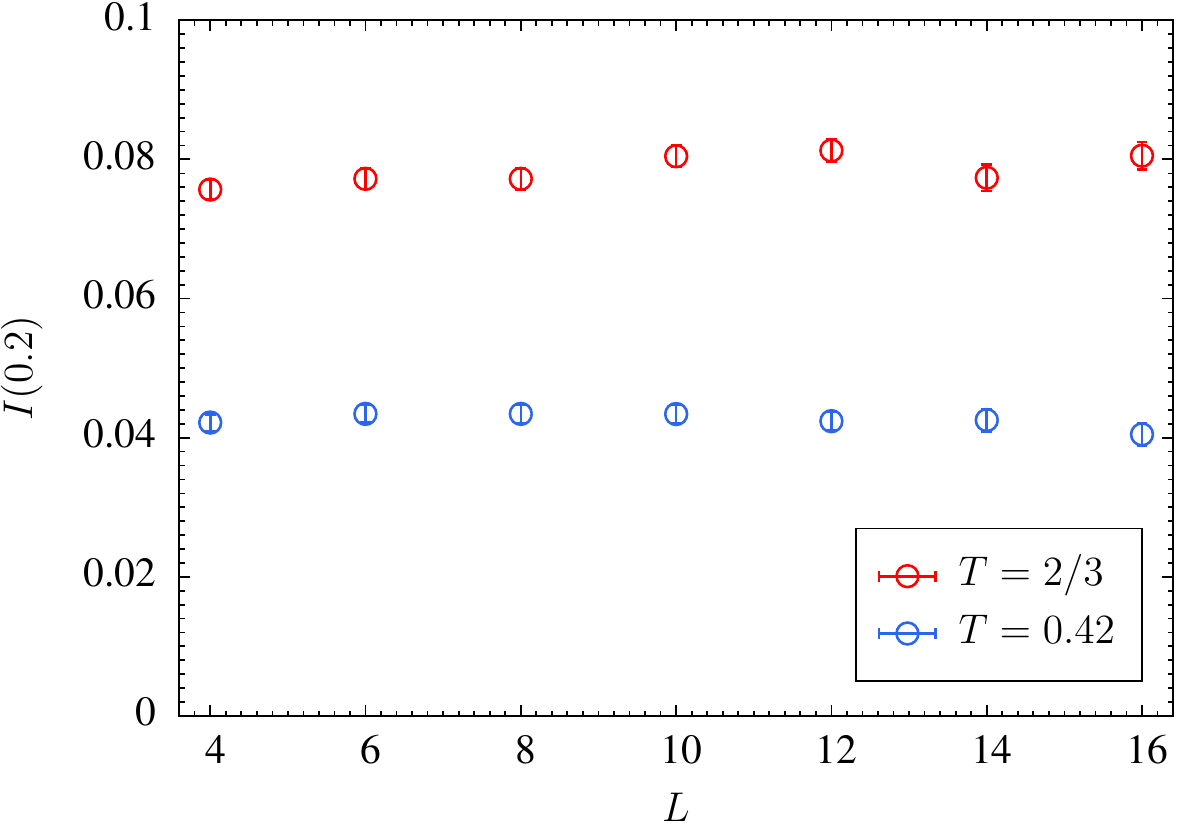}
\put(-204,155) {$(b)$} 
\caption{
\textit{Top panels:} The disorder-averaged overlap function for different system sizes $L$ at typical low temperatures $T=2/3$ (left) and $T=0.42$ (right), respectively.
\textit{Bottom panel:} The central weight $I(0.2)$ is approximately independent of the system size $L$ for both temperatures, in agreement with the many state picture. Note that the maximum size is $L=16$, compared with $L=12$ of \cite{Yucesoy:Delta}.
}
\label{Pq}
\end{center}
\end{figure}

\textit{Results--} The disorder-averaged overlap function and the central weight $I(0.2)$ at both $T=2/3$ and $0.42$ are shown in Fig.~\ref{Pq}. The central weight is essentially flat up to fluctuations as a function of the system size, in agreement with numerous previous results \cite{Enzo:Review,palassini:00,Yucesoy:Delta,Wang:PA}. 
This result is well known, except that we here further extend it to two larger system sizes at low temperatures. The result is in excellent agreement with RSB, but is strikingly different from the $1/L^{\theta_{\rm{DW}}}$ ($\theta_{\rm{DW}} \approx 0.2$) droplet scaling. To our best knowledge, there is no straightforward way to explain this as a finite-size effect, because the interface free energy scales well with this exponent for the same range of sizes. Finally, the weights of $T=0.42$ are smaller than those of $T=2/3$ as expected, as the effective number of ``active'' pure states is suppressed at lower temperatures.

\begin{figure*}[htb]
\begin{center}
\includegraphics[width=\columnwidth]{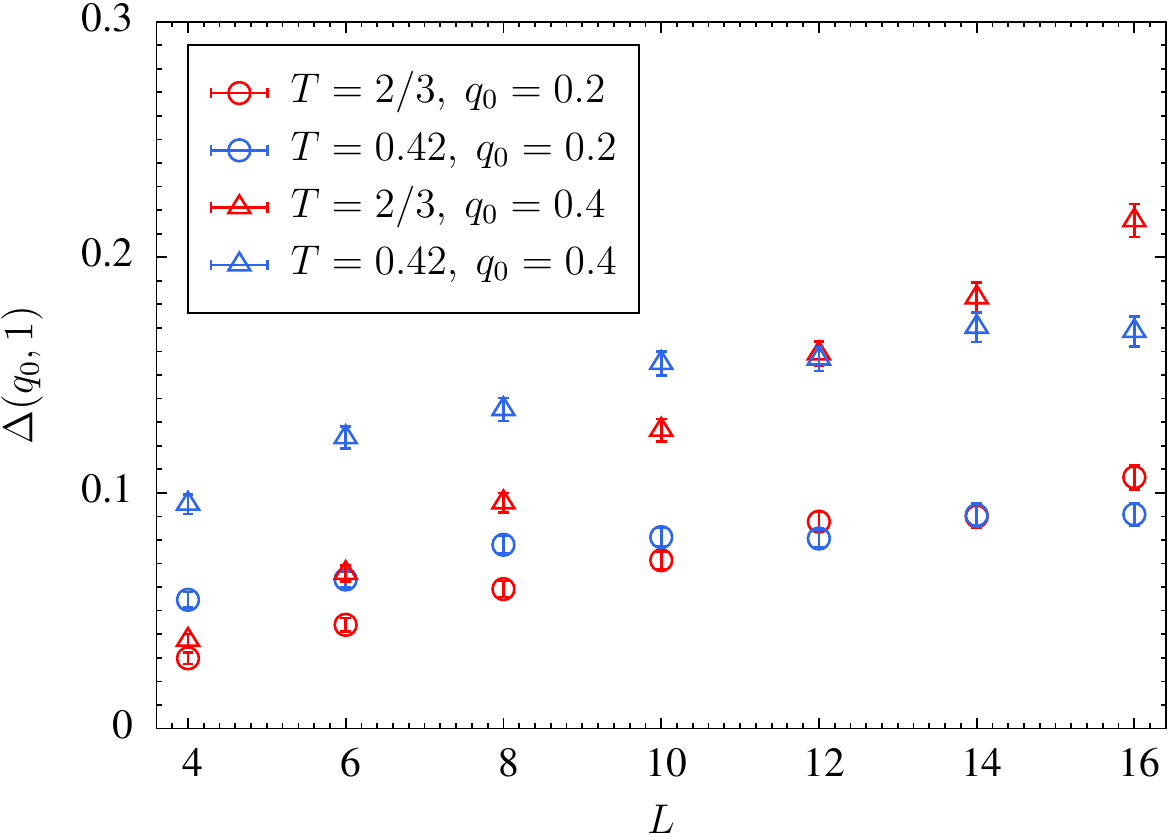}
\put (-18,158) {$(a)$} 
\includegraphics[width=\columnwidth]{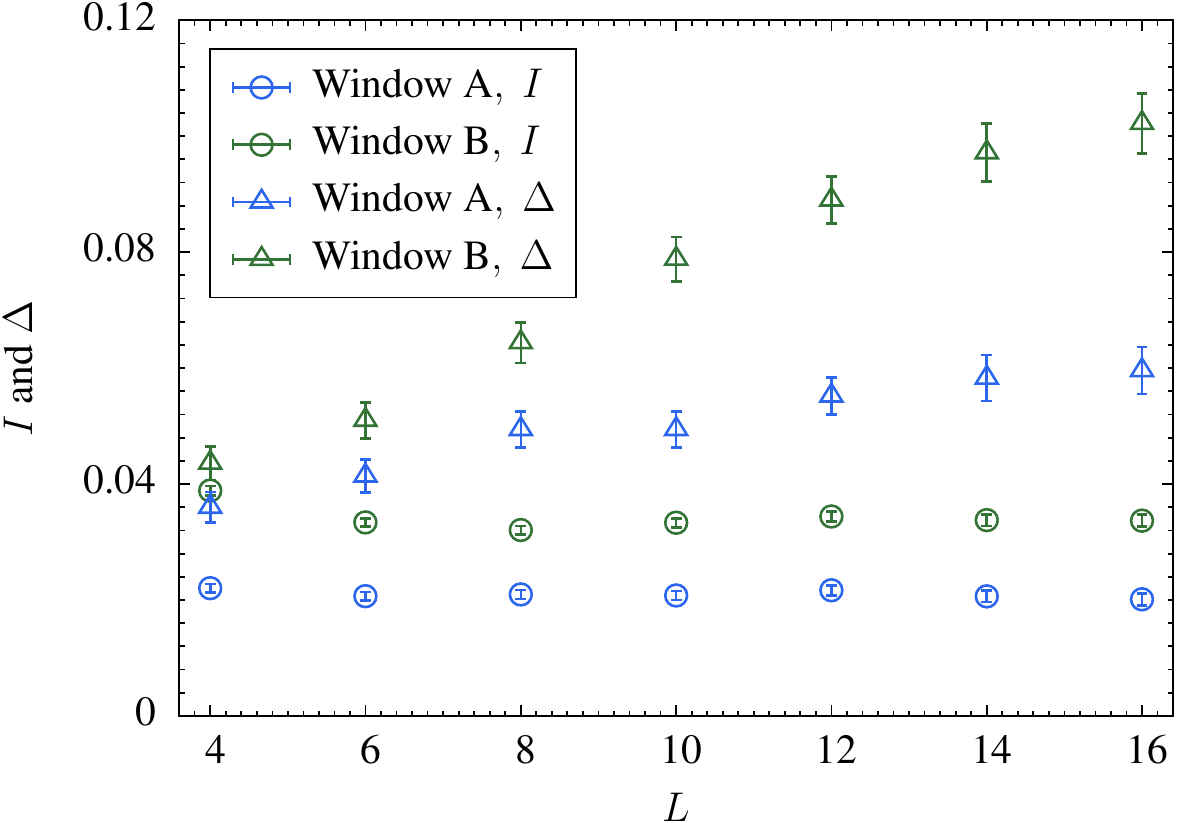}
\put (-18,156) {$(b)$} \\
\includegraphics[width=\columnwidth]{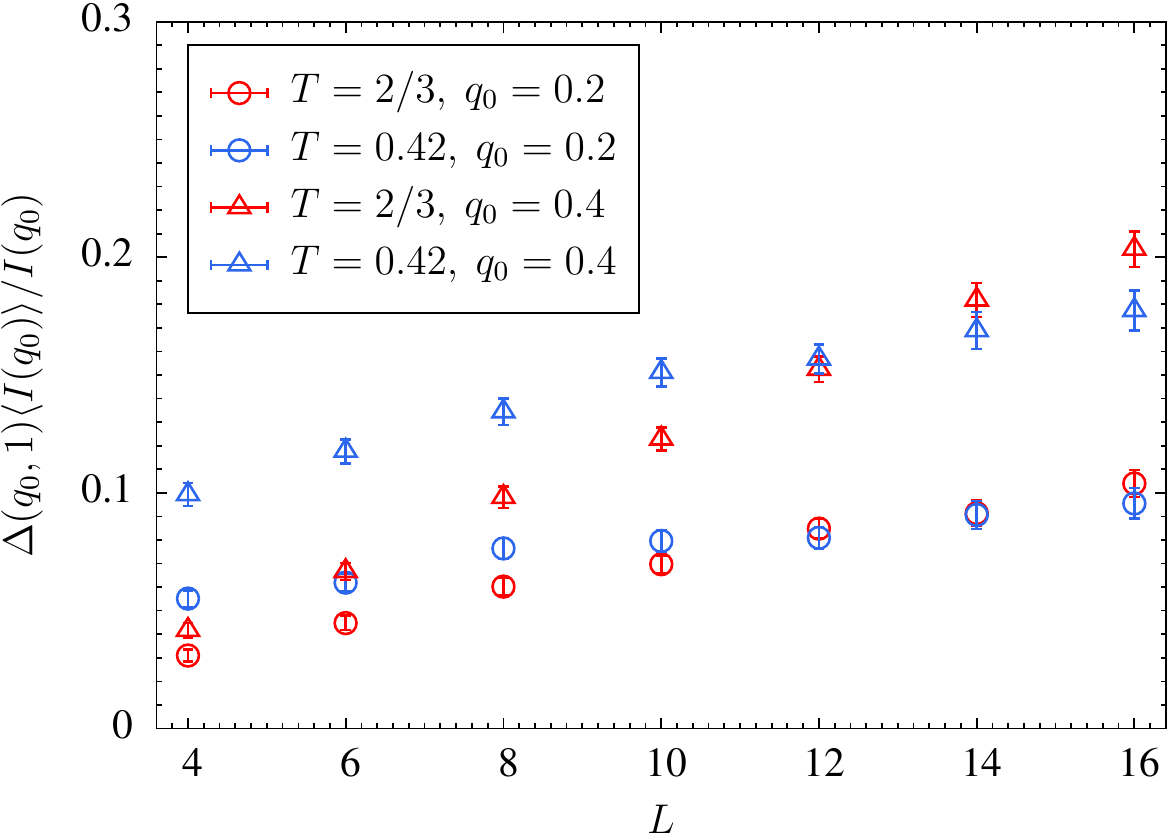}
\put (-18,158) {$(d)$} 
\includegraphics[width=\columnwidth]{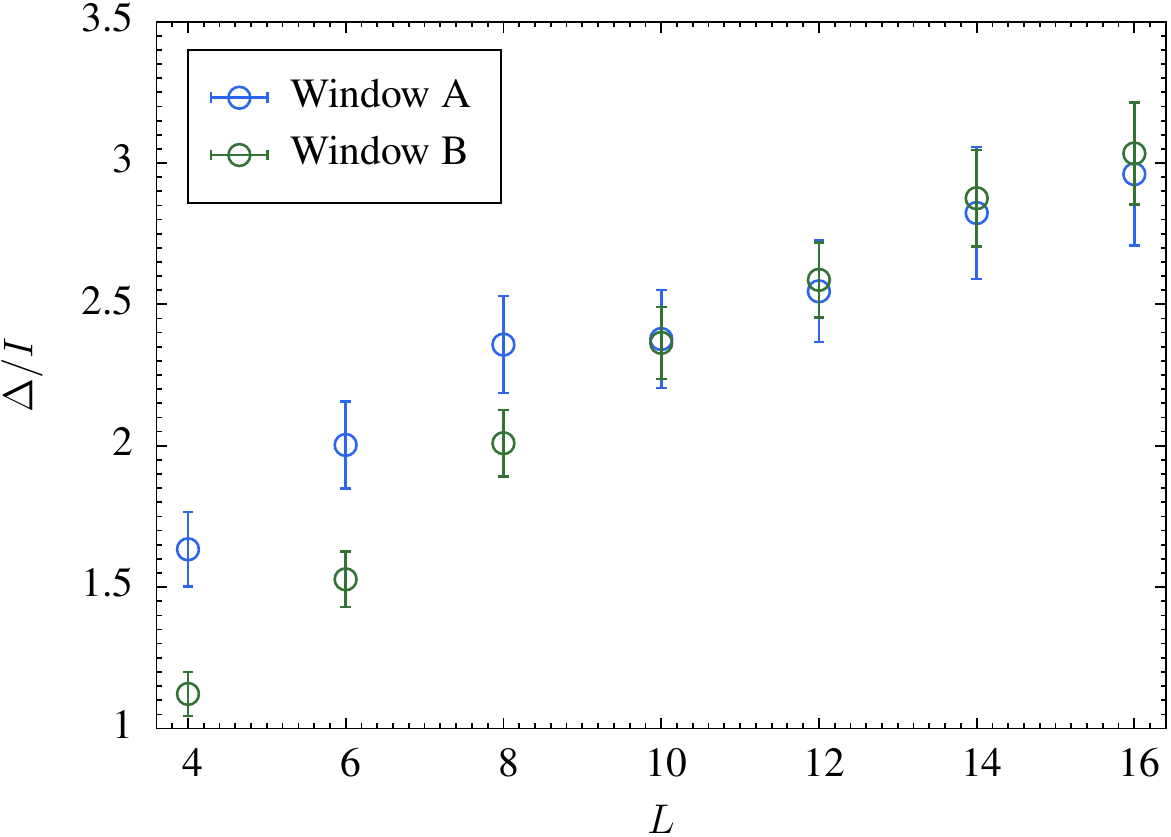}
\put (-18,158) {$(c)$}
\caption{
The statistics $\Delta(0.2,1)$ and $\Delta(0.4,1)$ as a function of $L$ at two temperatures $T=2/3$ and $T=0.42$ [panel $(a)$]. While $\Delta$ appears to have a growing trend, it is not fully regular. We find that $\Delta$ is subtly influenced by the central weight; cf. Fig.~\ref{Pq}. To illustrate this correlation, two windows of the same size but of very different weights are studied at $T=0.42$, and the Window B with a larger weight also has consistently larger $\Delta$ [panel $(b)$]. This correlation can be effectively reduced by looking at the statistic $\Delta/I$ as shown in the panel $(c)$. Similarly, the modified $\Delta$ as shown in panel $(d)$ has a much cleaner growth trend and is in fact remarkably monotonic. Here, $\langle I \rangle$ is the average of $I$ over the sizes. See the text for more details.
}
\label{Delta}
\end{center}
\end{figure*}

Next, we discuss the statistic $\Delta$ in detail. The results of $\Delta(0.2,1)$ and $\Delta(0.4,1)$ are shown in the top left panel of Fig.~\ref{Delta}. At first sight, the data look quite irregular as observed in \cite{Yucesoy:Delta}. There appears to be a growing trend in general, and the trend is much clearer at $T=2/3$ than at the lower temperature $T=0.42$. We shall discuss the effect of the temperature below, and first focus on the origin of the irregular low temperature data. We take the $\Delta(0.2,1)$ as an example without loss of generality. After an increasing trend at small sizes, the grow from $L=10$ to $12$ is very marginal (marginal or slightly negative depending on disorder realizations, here we call them collectively as marginal), in agreement with \cite{Yucesoy:Delta}. Then we observe a noticeable increase from $L=12$ to $14$, a somewhat promising result for many pure states. But the statistic subsequently grows rather marginally again from $L=14$ to $16$, resulting in a rather confusing situation. This puzzle is finally understood when we recognized a subtle correlation between $\Delta$ and the central weight. As illustrated in Fig.~\ref{Pq}, $I(0.2)$ drops slightly from $L=10$ to $12$, this is where the corresponding $\Delta$ has a marginal increase. Then $I(0.2)$ increases slightly from $L=12$ to $14$ and then drops slightly at $L=16$, explaining the correlated trend in $\Delta$. This together with other similar observations throughout our data collection process strongly suggest that the two statistics are correlated. Then $\Delta$ is presumably a growing function with increasing system size, but the growth is very sensitively affected by the fluctuations of the central weights, producing an irregular growth trend. This also partially explains why the data at a higher temperature have a clearer growth trend, as the relative fluctuation of the central weights with respect to the size-averaged mean is smaller at higher temperatures. An additional reason is that $\Delta$ has a larger growth rate at higher temperatures, note the crossings of the data at the two temperatures. 

The correlation between $\Delta$ and $I$ is reasonable from the following argument. The central weight is like a supply of peaks, and higher weights supply more peaks, which tend to statistically produce more peaked instances. Take two extreme examples, if $I(0.2)=0$, it is clear that $\Delta(0.2,1)$ is bounded to be zero. On the other hand, if $I_{\mathcal{J}}(0.2)$ for an instance is large, there is almost certainly a peak present, at least when the size is large. The more detailed sample-wise correlation is further illustrated in the Appendix. Moreover, it is found therein the correlation becomes stronger at larger sizes and lower temperatures, as the system moves closer to the $\delta$-peaks regime. This is reasonable as in this regime, the absence or presence of the central weight would directly correspond to $\Delta_{\mathcal{J}}=0$ or $1$, respectively. Since the fluctuation of $I$ has a profound influence on $\Delta$, we next look for modified statistics to compensate this correlation effect by a variance reduction method.


We define a slightly modified statistic of weighted $\Delta$ as $\Delta(q_0,\kappa)[\langle I(q_0) \rangle /I(q_0)]$, where the angular bracket is an average over the system size.  While this definition assumes in the first place $I$ is approximately constant, it has strong numerical supports as mentioned previously and it only slightly modifies the $\Delta$ data. Nevertheless, we shall present below another statistic which is similar in spirit but does not have this ``problem''. First, we look at an example to \textit{motivate} the weighted $\Delta$ using two windows that have very different weights, showing that it is effective for comparing $\Delta$ with different weights. From the overlap functions, it is clear that at the lower temperature there are wide $q$ ranges where the major $q_{\rm{EA}}$ peaks have little influence. In addition, the weight density is higher at larger $q$ than at the neighbourhood of $q=0$. Therefore, we select the following two windows and study the behaviours of $I$ and $\Delta$ at $T=0.42$: Window A as $q \in [-0.1, 0.1]$ with a small weight and Window B as $|q| \in [0.4, 0.5]$ with the same length but a noticeably larger weight. Here, $I$ and $\Delta$ are measured in the respective supports. The two statistics as a function of system size for these two windows are shown in the top right panel of Fig.~\ref{Delta}. Since Window B has consistently larger weights $I$, it also has consistently larger $\Delta$ as expected. The ratio $\Delta/I$ is shown in the bottom right panel, and this simple statistic brings the two $\Delta$ data sets much closer particularly for the pertinent large sizes. Similar behaviour is also found for the higher temperature, despite that Window B is slightly affected by the $q_{\rm{EA}}$ peaks. These demonstrate that $\Delta/I$ is an effective statistic to reduce the correlation effect from the weights, despite it may not fully remove it.

The modified $\Delta$ is shown in the bottom left panel of Fig.~\ref{Delta}. It is remarkable that this simple statistic has a clean growth trend, i.e., the growth trend is not only improved but also monotonic. We have carried out a quantitative growing trend test to leading linear order using a linear fit, as nearby data variations can be within error bars. The computed slopes are $0.0034(5)$ (low $T, q_0=0.2)$, $0.0062(4)$ (high $T, q_0=0.2)$, $0.0067(6)$ (low $T, q_0=0.4)$, $0.0138(5)$ (high $T, q_0=0.4)$, respectively. These values, especially the high temperature data, are all significantly larger than $0$, suggesting a collective growing trend of this statistic. We conclude therefore that the seemingly nonmonotonic growth of $\Delta$ is a result of its sensitivity to the fluctuations of the central weight. From this perspective, the statistic $\Delta$ is not as good as the central weight in discriminating the number of pure states due to its discrete nature as it amplifies fluctuations.

\begin{figure}[htb]
\begin{center}
\includegraphics[width=\columnwidth]{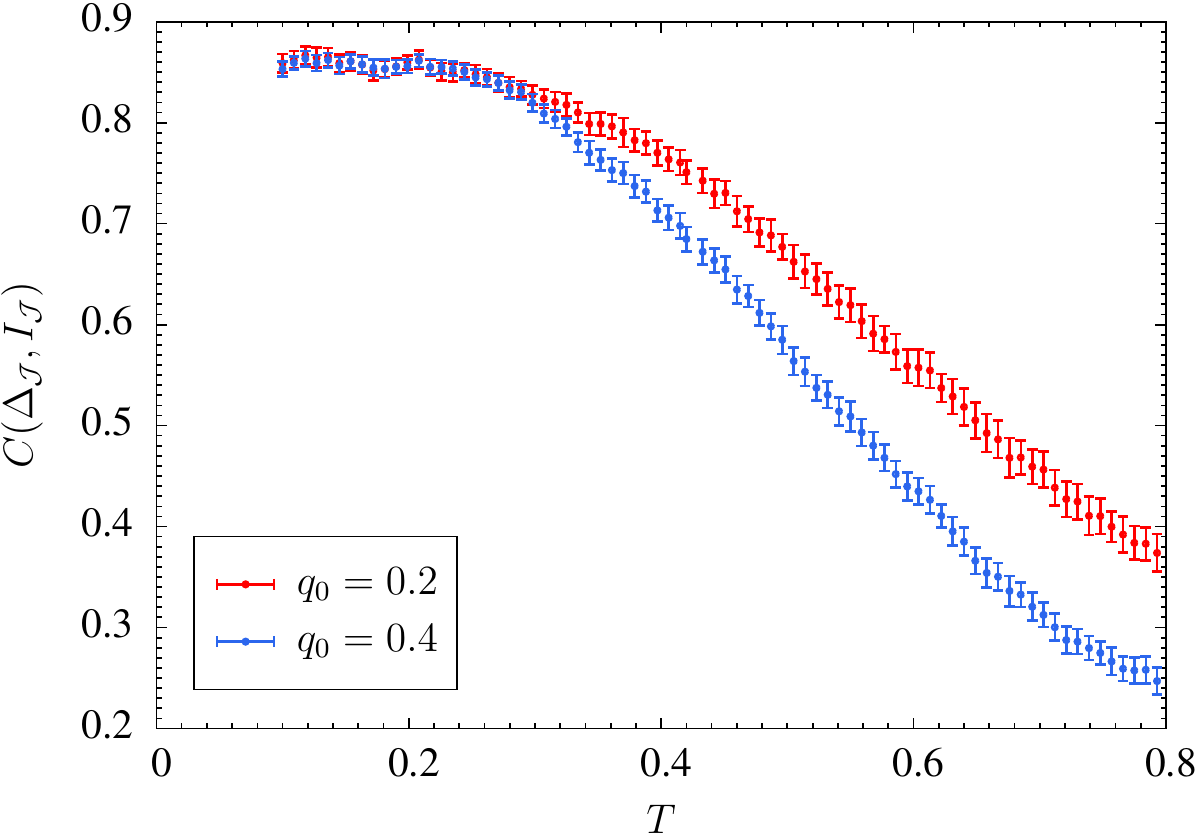}
\put (-211,157) {$(a)$} \\
\includegraphics[width=\columnwidth]{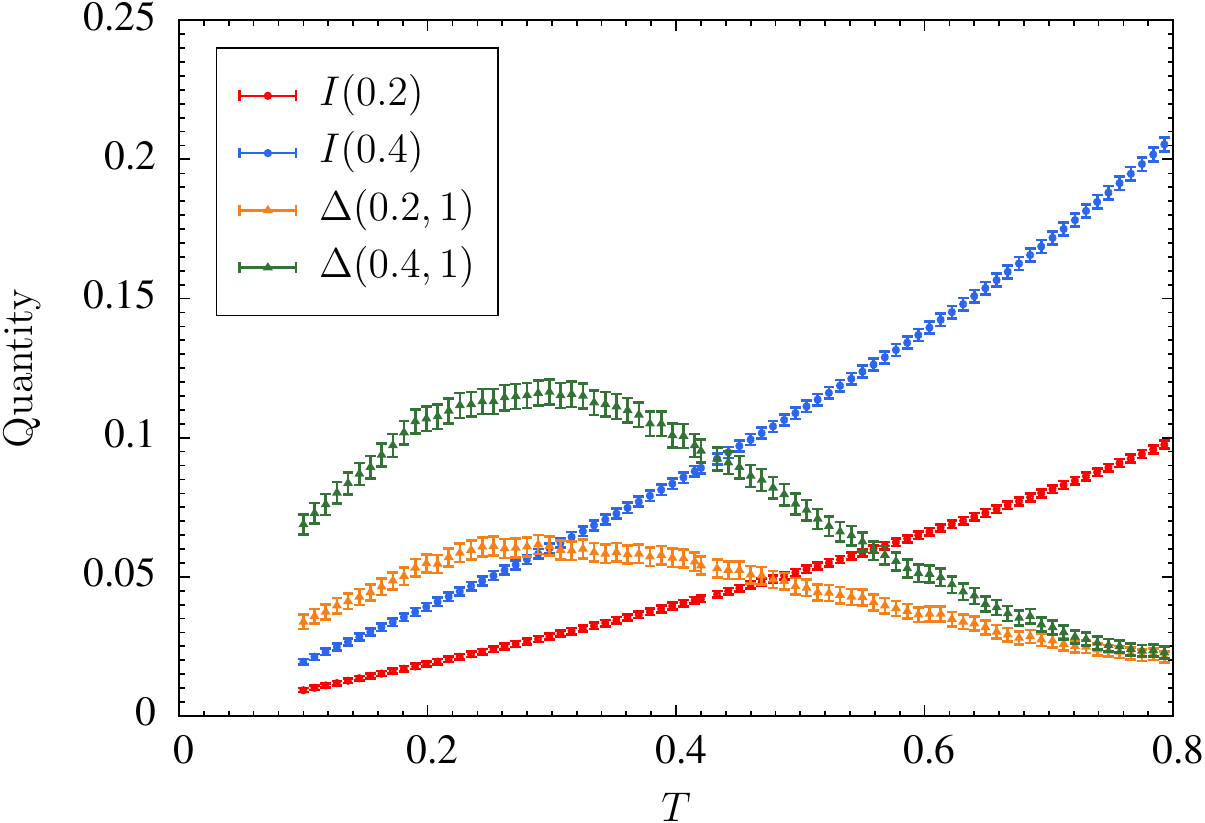}
\put(-206.5,27.5) {$(b)$} 
\caption{
Different from the large-size regime, the effect of the temperature on $\Delta$ for small sizes is complex because the rapidly increasing correlation and decreasing weight $I$ compete with each other as the temperature is decreased. The top panel $(a)$ shows the Pearson correlation coefficient and the bottom panel $(b)$ shows $\Delta$ and $I$ for $L=4$. Here, a local maximum of $\Delta$ is observed. As $T$ decreases, $\Delta$ first increases due to the rapidly growing correlation or peak sharpening between $0.3 \lesssim T <0.8$. When typical peaks are sufficiently narrow and tall $T \lesssim 0.3$, the weight effect takes over and $\Delta$ decreases with $I$, converging to $0$ as $T \rightarrow 0$. The figure also illustrates clearly the fluctuation of $\Delta$ is systematically larger than that of the central weight due to its discrete nature, and thus amplifies fluctuation.
}
\label{L4}
\end{center}
\end{figure}

We now comment on the effect of the temperature on the growth rate of $\Delta$. The modified $\Delta$ suggests that the growth rate at the higher temperature is noticeably larger than at the lower temperature. There is also an interesting crossing in $\Delta$ for each interval despite that $I$ has no crossing, showing the complex quantitative relations of $I$ and $\Delta$ in general. Nevertheless, the crossings can be qualitatively understood by the two peak-sharpening mechanisms mentioned above either by increasing the system size or lowering the temperature. When the system size is sufficiently large or the correlation is sufficiently strong, i.e., typical peaks regardless of the temperature are sharp and tall with respect to $\kappa$, the order of $I$ should sufficiently determine the order of $\Delta$ as the temperature is tuned. Then, the $\Delta$ at the higher temperature should be larger as the weight is larger. This is found numerically in Fig.~\ref{Delta}, as the red points exceed the blue points when the size is increased. On the other hand, the impact of the temperature on $\Delta$ for small sizes is, however, more complex. In this regime, correlation may strongly increase with decreasing temperature, i.e., typical peaks are wide and short at high temperatures and sharp and tall at low temperatures. Meanwhile, the weight decreases with decreasing temperature. Therefore, as the temperature decreases, these two effects \textit{compete} with each other. This provides a possibility that $\Delta$ may increase in a temperature interval as the temperature is decreased, and Fig.~\ref{Delta} suggests that this is actually realized, e.g., at $L=4$. Fortunately, it is not computationally hard to study the small size in detail in a wide range of temperatures. To this end, we have conducted an additional set of simulation to study the evolution of $\Delta(T)$ and $I(T)$ at $L=4$ down to $T=0.1$. Our picture is numerically confirmed, and the results are shown in Fig.~\ref{L4}. In a broad range of temperatures, here approximately $0.3 \lesssim T <0.8$, $\Delta$ grows with deceasing temperature. When typical peaks are sufficiently narrow and tall at low temperatures $T \lesssim 0.3$, similar to the large size regime, the weight effect takes over and $\Delta$ decreases along with decreasing $I$, converging to $0$ as $T \rightarrow 0$. The \textit{opposite} trends in the small and large size regimes strongly suggest that the growth rate of $\Delta$ is an increasing function of temperature in this wide temperature range $0.3 \lesssim T <0.8$, if we reasonably assume the growth is approximately linear particularly for the weighted $\Delta$. This is also in line with \cite{Yucesoy:Delta,billoire:13}, and our argument indicates that $\Delta$ should grow even slower and become more irregular at a slightly lower temperature, e.g., $T=0.3$.

\begin{figure}[htb]
\begin{center}
\includegraphics[width=\columnwidth]{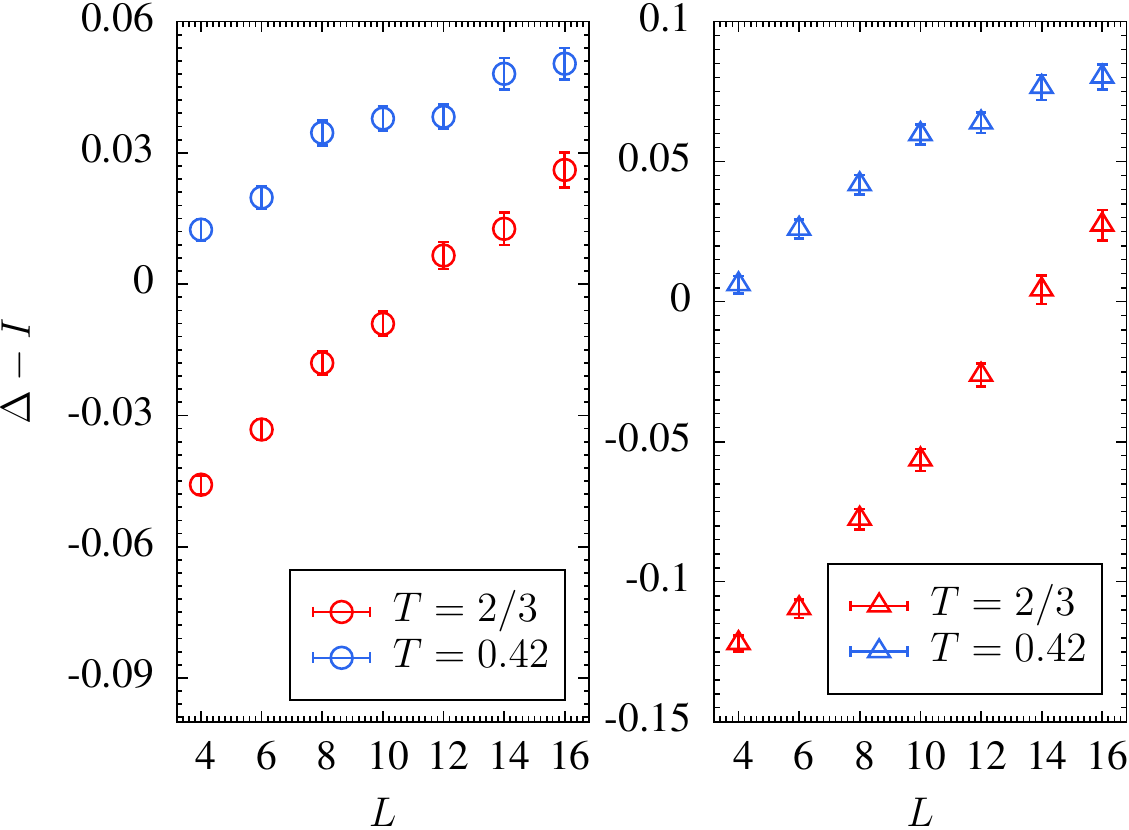}
\put (-204,29) {$(a)$} 
\put (-87,29) {$(b)$} \\
\includegraphics[width=\columnwidth]{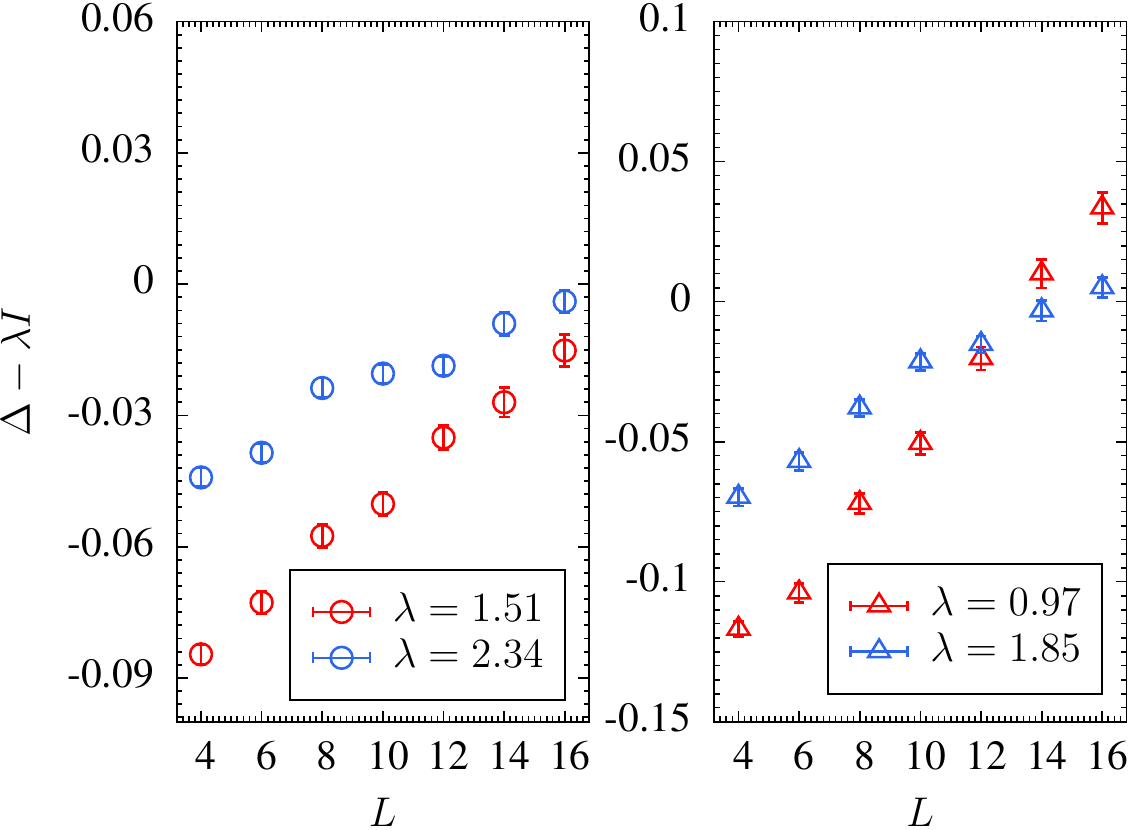}
\put (-132,166) {$(c)$}
\put (-15.5,166) {$(d)$}
\caption{
\textit{Top panels:} A more traditional statistic $\Delta-I$ to decorrelate $\Delta$ and $I$, and the left $(a)$ and right $(b)$ panels are for $q_0=0.2$ and $0.4$, respectively. The statistic shows a growing behaviour as the modified $\Delta$, in agreement with the many state picture. By contrast, the statistic should converge to $0$ for a two state picture. \textit{Bottom panels:} The same as above but for a refined statistic $\Delta-\lambda I$ using the control variate analysis with a size-averaged parameter $\lambda$. The data appear more regular as the errorbar is further suppressed, showing again a growing behaviour.
}
\label{Smats}
\end{center}
\end{figure}

Next we look at an alternative method for variance reduction \cite{HH:MC} by considering $\Delta-I$. The modified $\Delta$ above assumes $I$ is approximately constant, but this statistic does not need such an assumption. The results are shown in the top panels of Fig.~\ref{Smats}. Similar to the modified $\Delta$, this statistic also has a clean growth trend, in agreement with many pure states. By contrast, this statistic should converge to $0$ for a two state picture, as the two terms should separately converge to $0$. The data are clearly running above $0$ and are still growing, and should reach finite limits if there are many pure states. In addition, this statistic can be further refined using the control variate analysis \cite{Liu2004,Fernandez:MC}. This method utilizes a parametrized control variate estimator $\Delta - \lambda I$ where $\lambda$ is a free parameter. In order to minimize the variance of this statistic, it is straightforward to show that $\lambda=(\sigma_{\Delta}/\sigma_{I})C(\Delta_{\mathcal{J}}, I_{\mathcal{J}})$. The optimum $\lambda(T,q_0)$ depends on the temperature and the specific $\Delta$ window, and varies reasonably slowly with the system size. The size-averaged estimates are approximately $\lambda(2/3,0.2)=1.51, \lambda(0.42,0.2)=2.34, \lambda(2/3,0.4)=0.97$, and $\lambda(0.42,0.4)=1.85$, respectively. These refined statistics are shown in the bottom panels of Fig.~\ref{Smats}, showing similar behaviours, and it seems that the data are even more regular, as the errorbar is further suppressed particularly at the low temperature. Note that $\Delta -I$ remains important, despite its larger errorbar, as it clearly runs above $0$. Therefore, the statistics $I$, modified $\Delta$, $\Delta-I$ and $\Delta-\lambda I$ are all coherently in agreement with the many state picture.

Our results are in full qualitative agreement with \cite{Yucesoy:Delta,billoire:13}. The former group found a small growth rate of $\Delta$ at a low temperature, while the latter group found instead a much larger growth rate by operating at a higher temperature. The size of the central weight again has a significant impact on the growth behaviours of $\Delta$ \cite{Yucesoy:Delta}. Our large range of sizes is crucial in identifying this subtle correlation. Therefore, we interpret the low-temperature irregular $\Delta$ as fluctuations because of a small growth rate, rather than a genuine onset of the two state picture. We conclude that the statistic $\Delta$ has no evidence of violating many pure states, and instead it is consistent with a coherent many state picture \cite{billoire:13}. We cannot rule out the possibility that the two state picture behaviour is realized at yet much larger sizes currently not accessible, but we cannot glimpse such a trend, and it appears an unlikely scenario.

Finally, we briefly discuss the difficulties of the two state picture with current numerical results. (1) In the same range of sizes, we clearly get a finite domain-wall exponent $\theta_{\rm{DW}}$ yet a flat $I(0.2)$. It is unclear what finite-size effect is responsible for a flat central weight in the droplet framework. 
(2) The $\theta_{\rm{DW}}$ exponent is a growing function of dimensionality \cite{boettcher:05d} and even the SK model has a positive exponent \cite{Wang:KAS} which is clearly described by RSB. It seems likely that $\theta_{\rm{DW}}>0$ is not capable of excluding large-scale $O(1)$ droplet excitations as suggested by the droplet picture which assumes that droplet and domain-wall excitations are similar; see, e.g., \cite{Wang:EA4D} for an interesting possibility of how a positive domain-wall exponent and many pure states can coexist. Otherwise, an explanation should be provided on why this argument does not apply to, e.g., the SK model. These difficulties in our opinion must be addressed for a two state picture to be consistent.

\textit{Conclusions--} In this work we carried out a state-of-the-art simulation of the Gaussian Ising spin glass in three dimensions and examined the statistic $\Delta$ in detail. Our results reveal that the nonmonotonic growth of the statistic with system size is a result of its sensitivity to the fluctuations of the central weight $I$. By looking at a modified $\Delta$ and also $\Delta-I$ compensating for this correlation effect, we find essentially monotonic growth of the statistics. Combining with the relatively flat central weight, we conclude that the statistic $\Delta$ is in full agreement with the many state picture but not with the two state picture.

The spin-glass literature is overall currently far from conclusive, our investigation of the number of pure states benefits from running state-of-the-art simulations and using variance reduction data analysis. Since simulating much larger system sizes is currently not an option, it is highly motivated to carry out careful 
statistical analysis of the data to reduce the fluctuations in the observed quantities, as we have demonstrated here. 
To ultimately decide between different pictures might require considerably larger simulations that are presently out of reach, but our results suggest that the many pure state picture is most likely.


\section*{APPENDIX: Correlation between $\Delta_{\mathcal{J}}$ and $I_{\mathcal{J}}$}

\begin{figure}[htb]
\begin{center}
\includegraphics[width=\columnwidth]{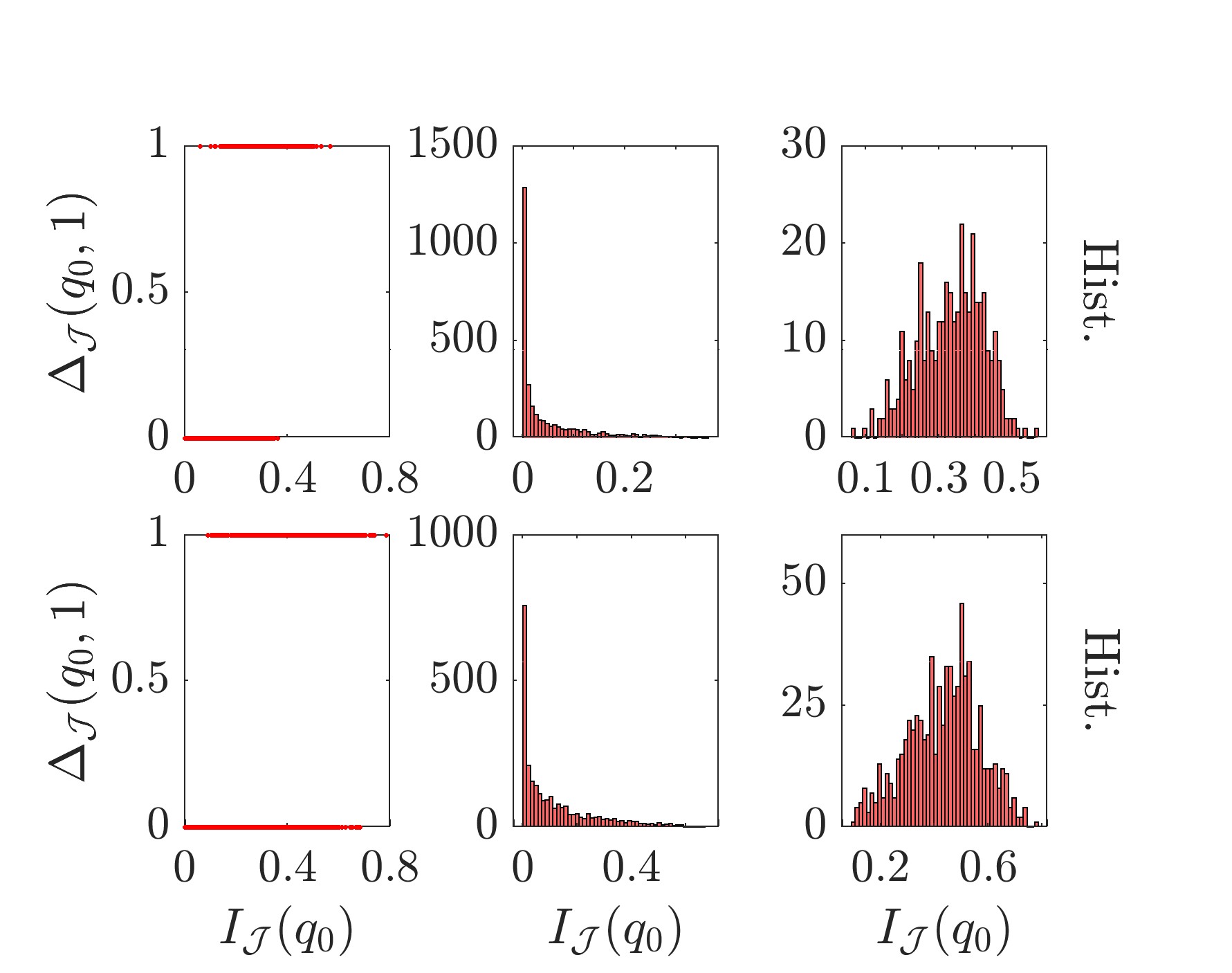}
\put (-181,110.7) {$(a)$}
\put (-180,33) {$(b)$} \\
\includegraphics[width=\columnwidth]{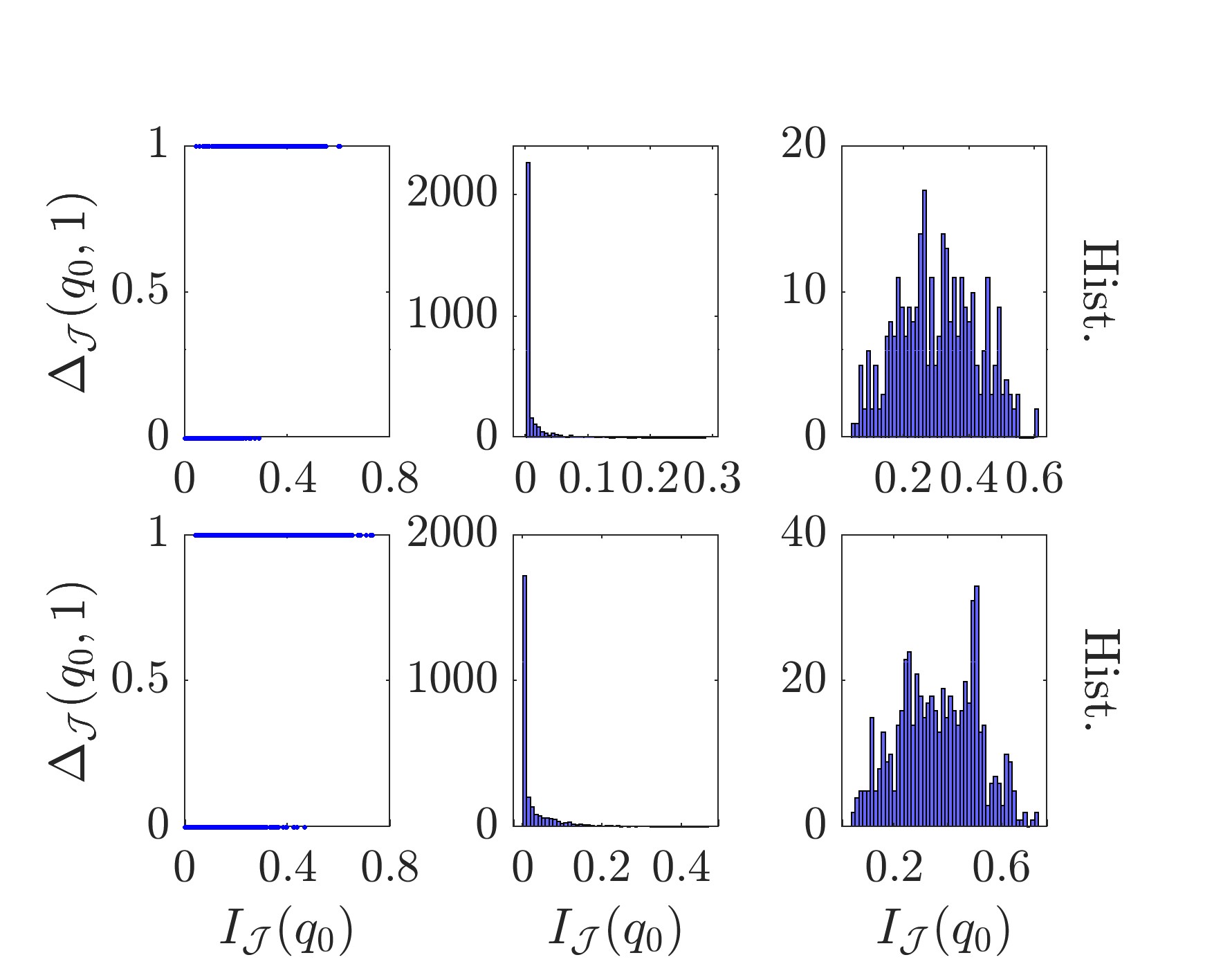}
\put (-180,110.7) {$(c)$}
\put (-181,33) {$(d)$} \\
\caption{
The left column of top panels $(a)$ shows the scatter plot of $\Delta_{\mathcal{J}}$ vs. $I_{\mathcal{J}}$ for $q_0=0.2$ at $T=2/3$. Note the empty top left and bottom right regions in the scatter plot, suggesting sample-wise correlation. To further illustrate the scatter plot, the histograms of $I_{\mathcal{J}}$ for the $\Delta_{\mathcal{J}}=0$ and $1$ classes are shown in the second and third columns, respectively. Similar correlations are found for $q_0=0.4$ [panels $(b)$], as well as the lower temperature $T=0.42$ in the corresponding panels $(c)$ and $(d)$. Other system sizes show very similar features.
}
\label{hist12}
\end{center}
\end{figure}

In this appendix, we characterize the correlation between $\Delta_{\mathcal{J}}$ and $I_{\mathcal{J}}$ in detail at the sample-wise level. The scatter plot of these two quantities of $L=16$ at both temperatures and intervals is shown in Fig.~\ref{hist12}. Since $\Delta$ is a discrete binary statistic, the conditional distributions of $I$ for $\Delta_{\mathcal{J}}=0$ and $1$ are illustrated. These plots, and also those of other system sizes, have very similar features.

\begin{table}
\caption{
Pearson correlation coefficient $C(\Delta_{\mathcal{J}}, I_{\mathcal{J}})$ for different system sizes $L$, temperatures $T$, and overlap intervals $[-q_0, q_0]$. Here, the pair represents $(T, q_0)$. The correlation is stronger at the lower temperature and larger sizes, where typical peaks are narrow and tall.
\label{table2}
}
\begin{tabular*}{\columnwidth}{@{\extracolsep{\fill}} l c c c r}
\hline
\hline
$L$  &$(2/3, 0.2)$  &$(2/3, 0.4)$ &$(0.42, 0.2)$ &$(0.42, 0.4)$ \\
\hline
$4$  &$0.485(17)$ &$0.351(14)$  &$0.751(13)$ &$0.685(14)$ \\
$6$  &$0.556(16)$ &$0.413(14)$  &$0.787(12)$ &$0.740(11)$ \\
$8$  &$0.629(15)$ &$0.508(13)$  &$0.808(10)$ &$0.772(09)$ \\
$10$ &$0.677(13)$ &$0.548(11)$  &$0.833(09)$ &$0.800(08)$ \\
$12$ &$0.717(11)$ &$0.616(10)$  &$0.838(09)$ &$0.805(08)$ \\
$14$ &$0.733(14)$ &$0.646(11)$  &$0.840(09)$ &$0.829(08)$ \\
$16$ &$0.751(12)$ &$0.670(10)$  &$0.861(08)$ &$0.832(08)$ \\
\hline
\hline
\end{tabular*}
\end{table} 

In the scatter plot, There is an interesting shift of $I_{\mathcal{J}}$ to larger values when $\Delta_{\mathcal{J}}$ steps from $0$ to $1$. This means that when the central weight is close to $0$, it is almost surely $\Delta_{\mathcal{J}}=0$. At the opposite end, samples of the largest central weights have almost surely $\Delta_{\mathcal{J}}=1$. In the histograms, the central weight conditional distribution has a very biased distribution towards $0$ for the class $\Delta_{\mathcal{J}}=0$, while it is approximately Gaussian distributed for the class $\Delta_{\mathcal{J}}=1$. Note that a sizeable fraction of samples have $I_{\mathcal{J}}$ close to $0$, indicating the correlation is prominent. The fraction of small weighted samples increases as the temperature is decreased, suggesting that the correlation should be stronger at lower temperatures.

In order to characterize the correlation more quantitatively, we have calculated the Pearson correlation coefficient and the results are summarized in Table~\ref{table2}. First, all coefficients are significantly larger than $0$, confirming the strong positive correlation between the two statistics. Moreover, the correlation becomes stronger at larger sizes and the lower temperature. A more detailed study of the size $L=4$ down to $T=0.1$ (see Fig.~\ref{L4}) suggests that the correlation is an increasing function of the decreasing temperature. It should be noted that in both cases the system is moving closer to the $\delta$-peaks regime. This trend is reasonable as in this regime, the absence or presence of the central weight would directly correspond to $\Delta_{\mathcal{J}}=0$ or $1$, respectively.



\textit{Acknowledgments--} We thank J. Machta, M. Weigel, and B. Yucesoy for helpful discussions. W.W.~acknowledges support from the Swedish Research Council Grant No.~642-2013-7837, and the Goran Gustafsson Foundation
for Research in Natural Sciences and Medicine, and the Fundamental Research Funds for the Central Universities, China. M.W.~acknowledges
support from the Swedish Research Council Grant No.~621-2012-3984.
The computations were performed on resources provided by the Swedish
National Infrastructure for Computing (SNIC) at the National
Supercomputer Centre (NSC), and the High Performance Computing Center
North (HPC2N), and the Emei cluster at Sichuan university.

\bibliography{Refs}

\begin{thebibliography}{46}
\expandafter\ifx\csname natexlab\endcsname\relax\def\natexlab#1{#1}\fi
\expandafter\ifx\csname bibnamefont\endcsname\relax
  \def\bibnamefont#1{#1}\fi
\expandafter\ifx\csname bibfnamefont\endcsname\relax
  \def\bibfnamefont#1{#1}\fi
\expandafter\ifx\csname citenamefont\endcsname\relax
  \def\citenamefont#1{#1}\fi
\expandafter\ifx\csname url\endcsname\relax
  \def\url#1{\texttt{#1}}\fi
\expandafter\ifx\csname urlprefix\endcsname\relax\def\urlprefix{URL }\fi
\providecommand{\bibinfo}[2]{#2}
\providecommand{\eprint}[2][]{\url{#2}}

\bibitem[{\citenamefont{Binder and Young}(1986)}]{Young:RMP}
\bibinfo{author}{\bibfnamefont{K.}~\bibnamefont{Binder}} \bibnamefont{and}
  \bibinfo{author}{\bibfnamefont{A.~P.} \bibnamefont{Young}},
  \emph{\bibinfo{title}{{Spin Glasses: Experimental Facts, Theoretical Concepts
  and Open Questions}}}, \bibinfo{journal}{Rev. Mod. Phys.}
  \textbf{\bibinfo{volume}{58}}, \bibinfo{pages}{801} (\bibinfo{year}{1986}).

\bibitem[{\citenamefont{Stein and Newman}(2013)}]{book}
\bibinfo{author}{\bibfnamefont{D.}~\bibnamefont{Stein}} \bibnamefont{and}
  \bibinfo{author}{\bibfnamefont{C.}~\bibnamefont{Newman}},
  \emph{\bibinfo{title}{Spin Glasses and Complexity}}, Primers in Complex
  Systems (\bibinfo{publisher}{Princeton University Press},
  \bibinfo{year}{2013}).

\bibitem[{\citenamefont{Sherrington and Kirkpatrick}(1975)}]{SK}
\bibinfo{author}{\bibfnamefont{D.}~\bibnamefont{Sherrington}} \bibnamefont{and}
  \bibinfo{author}{\bibfnamefont{S.}~\bibnamefont{Kirkpatrick}},
  \emph{\bibinfo{title}{Solvable model of a spin glass}},
  \bibinfo{journal}{Phys. Rev. Lett.} \textbf{\bibinfo{volume}{35}},
  \bibinfo{pages}{1792} (\bibinfo{year}{1975}).

\bibitem[{\citenamefont{Parisi}(1979)}]{parisi:79}
\bibinfo{author}{\bibfnamefont{G.}~\bibnamefont{Parisi}},
  \emph{\bibinfo{title}{Infinite number of order parameters for spin-glasses}},
  \bibinfo{journal}{Phys. Rev. Lett.} \textbf{\bibinfo{volume}{43}},
  \bibinfo{pages}{1754} (\bibinfo{year}{1979}).

\bibitem[{\citenamefont{Parisi}(1980)}]{parisi:80}
\bibinfo{author}{\bibfnamefont{G.}~\bibnamefont{Parisi}},
  \emph{\bibinfo{title}{The order parameter for spin glasses: a function on the
  interval $0$--$1$}}, \bibinfo{journal}{J. Phys. A}
  \textbf{\bibinfo{volume}{13}}, \bibinfo{pages}{1101} (\bibinfo{year}{1980}).

\bibitem[{\citenamefont{Parisi}(1983)}]{parisi:83}
\bibinfo{author}{\bibfnamefont{G.}~\bibnamefont{Parisi}},
  \emph{\bibinfo{title}{Order parameter for spin-glasses}},
  \bibinfo{journal}{Phys. Rev. Lett.} \textbf{\bibinfo{volume}{50}},
  \bibinfo{pages}{1946} (\bibinfo{year}{1983}).

\bibitem[{\citenamefont{Edwards and Anderson}(1975)}]{EA}
\bibinfo{author}{\bibfnamefont{S.~F.} \bibnamefont{Edwards}} \bibnamefont{and}
  \bibinfo{author}{\bibfnamefont{P.~W.} \bibnamefont{Anderson}},
  \emph{\bibinfo{title}{Theory of spin glasses}}, \bibinfo{journal}{J. Phys. F:
  Met. Phys.} \textbf{\bibinfo{volume}{5}}, \bibinfo{pages}{965}
  (\bibinfo{year}{1975}).

\bibitem[{\citenamefont{Parisi}(2008)}]{parisi:08}
\bibinfo{author}{\bibfnamefont{G.}~\bibnamefont{Parisi}},
  \emph{\bibinfo{title}{{{Some considerations of finite dimensional spin
  glasses}}}}, \bibinfo{journal}{J. Phys. A} \textbf{\bibinfo{volume}{41}},
  \bibinfo{pages}{324002} (\bibinfo{year}{2008}).

\bibitem[{\citenamefont{M\'ezard et~al.}(1987)\citenamefont{M\'ezard, Parisi,
  and Virasoro}}]{mezard:87}
\bibinfo{author}{\bibfnamefont{M.}~\bibnamefont{M\'ezard}},
  \bibinfo{author}{\bibfnamefont{G.}~\bibnamefont{Parisi}}, \bibnamefont{and}
  \bibinfo{author}{\bibfnamefont{M.~A.} \bibnamefont{Virasoro}},
  \emph{\bibinfo{title}{Spin Glass Theory and Beyond}}
  (\bibinfo{publisher}{World Scientific}, \bibinfo{address}{Singapore},
  \bibinfo{year}{1987}).

\bibitem[{\citenamefont{McMillan}(1984)}]{mcmillan:84b}
\bibinfo{author}{\bibfnamefont{W.~L.} \bibnamefont{McMillan}},
  \emph{\bibinfo{title}{Domain-wall renormalization-group study of the
  two-dimensional random {I}sing model}}, \bibinfo{journal}{Phys. Rev. B}
  \textbf{\bibinfo{volume}{29}}, \bibinfo{pages}{4026} (\bibinfo{year}{1984}).

\bibitem[{\citenamefont{Bray and Moore}(1986)}]{bray:86}
\bibinfo{author}{\bibfnamefont{A.~J.} \bibnamefont{Bray}} \bibnamefont{and}
  \bibinfo{author}{\bibfnamefont{M.~A.} \bibnamefont{Moore}},
  \emph{\bibinfo{title}{Scaling theory of the ordered phase of spin glasses}},
  in \emph{\bibinfo{booktitle}{Heidelberg Colloquium on Glassy Dynamics and
  Optimization}}, edited by
  \bibinfo{editor}{\bibfnamefont{L.}~\bibnamefont{Van~Hemmen}}
  \bibnamefont{and}
  \bibinfo{editor}{\bibfnamefont{I.}~\bibnamefont{Morgenstern}}
  (\bibinfo{publisher}{Springer}, \bibinfo{address}{New York},
  \bibinfo{year}{1986}), p. \bibinfo{pages}{121}.

\bibitem[{\citenamefont{Fisher and Huse}(1986)}]{fisher:86}
\bibinfo{author}{\bibfnamefont{D.~S.} \bibnamefont{Fisher}} \bibnamefont{and}
  \bibinfo{author}{\bibfnamefont{D.~A.} \bibnamefont{Huse}},
  \emph{\bibinfo{title}{Ordered phase of short-range {I}sing spin-glasses}},
  \bibinfo{journal}{Phys. Rev. Lett.} \textbf{\bibinfo{volume}{56}},
  \bibinfo{pages}{1601} (\bibinfo{year}{1986}).

\bibitem[{\citenamefont{Fisher and Huse}(1987)}]{fisher:87}
\bibinfo{author}{\bibfnamefont{D.~S.} \bibnamefont{Fisher}} \bibnamefont{and}
  \bibinfo{author}{\bibfnamefont{D.~A.} \bibnamefont{Huse}},
  \emph{\bibinfo{title}{Absence of many states in realistic spin glasses}},
  \bibinfo{journal}{J. Phys. A} \textbf{\bibinfo{volume}{20}},
  \bibinfo{pages}{L1005} (\bibinfo{year}{1987}).

\bibitem[{\citenamefont{Fisher and Huse}(1988)}]{fisher:88}
\bibinfo{author}{\bibfnamefont{D.~S.} \bibnamefont{Fisher}} \bibnamefont{and}
  \bibinfo{author}{\bibfnamefont{D.~A.} \bibnamefont{Huse}},
  \emph{\bibinfo{title}{Equilibrium behavior of the spin-glass ordered phase}},
  \bibinfo{journal}{Phys. Rev. B} \textbf{\bibinfo{volume}{38}},
  \bibinfo{pages}{386} (\bibinfo{year}{1988}).

\bibitem[{\citenamefont{{Wang} et~al.}(2017)\citenamefont{{Wang}, {Moore}, and
  {Katzgraber}}}]{Wang:Fractal}
\bibinfo{author}{\bibfnamefont{W.}~\bibnamefont{{Wang}}},
  \bibinfo{author}{\bibfnamefont{M.~A.} \bibnamefont{{Moore}}},
  \bibnamefont{and} \bibinfo{author}{\bibfnamefont{H.~G.}
  \bibnamefont{{Katzgraber}}}, \emph{\bibinfo{title}{{{Fractal Dimension of
  Interfaces in Edwards-Anderson and Long-range Ising Spin Glasses: Determining
  the Applicability of Different Theoretical Descriptions}}}},
  \bibinfo{journal}{Phys. Rev. Lett.} \textbf{\bibinfo{volume}{119}},
  \bibinfo{pages}{100602} (\bibinfo{year}{2017}).

\bibitem[{\citenamefont{de~Almeida and Thouless}(1978)}]{almeida:78}
\bibinfo{author}{\bibfnamefont{J.~R.~L.} \bibnamefont{de~Almeida}}
  \bibnamefont{and} \bibinfo{author}{\bibfnamefont{D.~J.}
  \bibnamefont{Thouless}}, \emph{\bibinfo{title}{Stability of the
  {S}herrington-{K}irkpatrick solution of a spin glass model}},
  \bibinfo{journal}{J. Phys. A} \textbf{\bibinfo{volume}{11}},
  \bibinfo{pages}{983} (\bibinfo{year}{1978}).

\bibitem[{\citenamefont{Charbonneau et~al.}(2014)\citenamefont{Charbonneau,
  Kurchan, Parisi, Urbani, and Zamponi}}]{charbonneau:14}
\bibinfo{author}{\bibfnamefont{P.}~\bibnamefont{Charbonneau}},
  \bibinfo{author}{\bibfnamefont{J.}~\bibnamefont{Kurchan}},
  \bibinfo{author}{\bibfnamefont{G.}~\bibnamefont{Parisi}},
  \bibinfo{author}{\bibfnamefont{P.}~\bibnamefont{Urbani}}, \bibnamefont{and}
  \bibinfo{author}{\bibfnamefont{F.}~\bibnamefont{Zamponi}},
  \emph{\bibinfo{title}{{{Fractal free energy landscapes in structural
  glasses}}}}, \bibinfo{journal}{Nat. Comm.} \textbf{\bibinfo{volume}{5}},
  \bibinfo{pages}{3725} (\bibinfo{year}{2014}).

\bibitem[{\citenamefont{Hicks et~al.}(2018)\citenamefont{Hicks, Wheatley,
  Godfrey, and Moore}}]{Mike:GT}
\bibinfo{author}{\bibfnamefont{C.~L.} \bibnamefont{Hicks}},
  \bibinfo{author}{\bibfnamefont{M.~J.} \bibnamefont{Wheatley}},
  \bibinfo{author}{\bibfnamefont{M.~J.} \bibnamefont{Godfrey}},
  \bibnamefont{and} \bibinfo{author}{\bibfnamefont{M.~A.} \bibnamefont{Moore}},
  \emph{\bibinfo{title}{{Gardner Transition in Physical Dimensions}}},
  \bibinfo{journal}{Phys. Rev. Lett.} \textbf{\bibinfo{volume}{120}},
  \bibinfo{pages}{225501} (\bibinfo{year}{2018}).

\bibitem[{\citenamefont{Palassini and Young}(1999)}]{Young:GS}
\bibinfo{author}{\bibfnamefont{M.}~\bibnamefont{Palassini}} \bibnamefont{and}
  \bibinfo{author}{\bibfnamefont{A.~P.} \bibnamefont{Young}},
  \emph{\bibinfo{title}{{Triviality of the Ground State Structure in {I}sing
  Spin Glasses}}}, \bibinfo{journal}{Phys. Rev. Lett.}
  \textbf{\bibinfo{volume}{83}}, \bibinfo{pages}{5126} (\bibinfo{year}{1999}).

\bibitem[{\citenamefont{Hatano and Gubernatis}(2002)}]{Hatano:GS}
\bibinfo{author}{\bibfnamefont{N.}~\bibnamefont{Hatano}} \bibnamefont{and}
  \bibinfo{author}{\bibfnamefont{J.~E.} \bibnamefont{Gubernatis}},
  \emph{\bibinfo{title}{{Evidence for the double degeneracy of the ground state
  in the three-dimensional $\ifmmode\pm\else\textpm\fi{}J$ spin glass}}},
  \bibinfo{journal}{Phys. Rev. B} \textbf{\bibinfo{volume}{66}},
  \bibinfo{pages}{054437} (\bibinfo{year}{2002}).

\bibitem[{\citenamefont{{Alvarez Ba{\~n}os} et~al.}(2010)\citenamefont{{Alvarez
  Ba{\~n}os}, {Cruz}, {Fernandez}, {Gil-Narvion}, {Gordillo-Guerrero},
  {Guidetti}, {Maiorano}, {Mantovani}, {Marinari}, {Martin-Mayor}
  et~al.}}]{alvarez:10a}
\bibinfo{author}{\bibfnamefont{R.}~\bibnamefont{{Alvarez Ba{\~n}os}}},
  \bibinfo{author}{\bibfnamefont{A.}~\bibnamefont{{Cruz}}},
  \bibinfo{author}{\bibfnamefont{L.~A.} \bibnamefont{{Fernandez}}},
  \bibinfo{author}{\bibfnamefont{J.~M.} \bibnamefont{{Gil-Narvion}}},
  \bibinfo{author}{\bibfnamefont{A.}~\bibnamefont{{Gordillo-Guerrero}}},
  \bibinfo{author}{\bibfnamefont{M.}~\bibnamefont{{Guidetti}}},
  \bibinfo{author}{\bibfnamefont{A.}~\bibnamefont{{Maiorano}}},
  \bibinfo{author}{\bibfnamefont{F.}~\bibnamefont{{Mantovani}}},
  \bibinfo{author}{\bibfnamefont{E.}~\bibnamefont{{Marinari}}},
  \bibinfo{author}{\bibfnamefont{V.}~\bibnamefont{{Martin-Mayor}}},
  \bibnamefont{et~al.}, \emph{\bibinfo{title}{{Nature of the spin-glass phase
  at experimental length scales}}}, \bibinfo{journal}{J. Stat. Mech.}
  \textbf{\bibinfo{volume}{\normalfont{P06026}}} (\bibinfo{year}{2010}).

\bibitem[{\citenamefont{Katzgraber and Young}(2002)}]{Katzgraber:FBC}
\bibinfo{author}{\bibfnamefont{H.~G.} \bibnamefont{Katzgraber}}
  \bibnamefont{and} \bibinfo{author}{\bibfnamefont{A.~P.} \bibnamefont{Young}},
  \emph{\bibinfo{title}{{M}onte {C}arlo simulations of spin-glasses at low
  temperatures: {E}ffects of free boundary conditions}},
  \bibinfo{journal}{Phys. Rev. B} \textbf{\bibinfo{volume}{65}},
  \bibinfo{pages}{214402} (\bibinfo{year}{2002}).

\bibitem[{\citenamefont{Wang}(2017)}]{Wang:FBC}
\bibinfo{author}{\bibfnamefont{W.}~\bibnamefont{Wang}},
  \emph{\bibinfo{title}{{Numerical simulations of Ising spin glasses with free
  boundary conditions: The role of droplet excitations and domain walls}}},
  \bibinfo{journal}{Phys. Rev. E} \textbf{\bibinfo{volume}{95}},
  \bibinfo{pages}{032143} (\bibinfo{year}{2017}).

\bibitem[{\citenamefont{Wittmann et~al.}(2014)\citenamefont{Wittmann, Yucesoy,
  Katzgraber, Machta, and Young}}]{Yucesoy:Delta2}
\bibinfo{author}{\bibfnamefont{M.}~\bibnamefont{Wittmann}},
  \bibinfo{author}{\bibfnamefont{B.}~\bibnamefont{Yucesoy}},
  \bibinfo{author}{\bibfnamefont{H.~G.} \bibnamefont{Katzgraber}},
  \bibinfo{author}{\bibfnamefont{J.}~\bibnamefont{Machta}}, \bibnamefont{and}
  \bibinfo{author}{\bibfnamefont{A.~P.} \bibnamefont{Young}},
  \emph{\bibinfo{title}{{Low-temperature behavior of the statistics of the
  overlap distribution in Ising spin-glass models}}}, \bibinfo{journal}{Phys.
  Rev. B} \textbf{\bibinfo{volume}{90}}, \bibinfo{pages}{134419}
  (\bibinfo{year}{2014}).

\bibitem[{\citenamefont{Yucesoy et~al.}(2012)\citenamefont{Yucesoy, Katzgraber,
  and Machta}}]{Yucesoy:Delta}
\bibinfo{author}{\bibfnamefont{B.}~\bibnamefont{Yucesoy}},
  \bibinfo{author}{\bibfnamefont{H.~G.} \bibnamefont{Katzgraber}},
  \bibnamefont{and} \bibinfo{author}{\bibfnamefont{J.}~\bibnamefont{Machta}},
  \emph{\bibinfo{title}{{Evidence of Non-Mean-Field-Like Low-Temperature
  Behavior in the Edwards-Anderson Spin-Glass Model}}}, \bibinfo{journal}{Phys.
  Rev. Lett.} \textbf{\bibinfo{volume}{109}}, \bibinfo{pages}{177204}
  (\bibinfo{year}{2012}).

\bibitem[{\citenamefont{Billoire et~al.}(2013)\citenamefont{Billoire,
  Fernandez, Maiorano, Marinari, Martin-Mayor, Parisi, Ricci-Tersenghi,
  Ruiz-Lorenzo, and Yllanes}}]{billoire:13}
\bibinfo{author}{\bibfnamefont{A.}~\bibnamefont{Billoire}},
  \bibinfo{author}{\bibfnamefont{L.~A.} \bibnamefont{Fernandez}},
  \bibinfo{author}{\bibfnamefont{A.}~\bibnamefont{Maiorano}},
  \bibinfo{author}{\bibfnamefont{E.}~\bibnamefont{Marinari}},
  \bibinfo{author}{\bibfnamefont{V.}~\bibnamefont{Martin-Mayor}},
  \bibinfo{author}{\bibfnamefont{G.}~\bibnamefont{Parisi}},
  \bibinfo{author}{\bibfnamefont{F.}~\bibnamefont{Ricci-Tersenghi}},
  \bibinfo{author}{\bibfnamefont{J.~J.} \bibnamefont{Ruiz-Lorenzo}},
  \bibnamefont{and} \bibinfo{author}{\bibfnamefont{D.}~\bibnamefont{Yllanes}},
  \emph{\bibinfo{title}{{{Comment on ''Evidence of Non-Mean-Field-Like
  Low-Temperature Behavior in the Edwards-Anderson Spin-Glass Model''}}}},
  \bibinfo{journal}{Phys. Rev. Lett.} \textbf{\bibinfo{volume}{110}},
  \bibinfo{pages}{219701} (\bibinfo{year}{2013}).

\bibitem[{\citenamefont{Yucesoy et~al.}(2013)\citenamefont{Yucesoy, Katzgraber,
  and Machta}}]{Yucesoy:Reply}
\bibinfo{author}{\bibfnamefont{B.}~\bibnamefont{Yucesoy}},
  \bibinfo{author}{\bibfnamefont{H.~G.} \bibnamefont{Katzgraber}},
  \bibnamefont{and} \bibinfo{author}{\bibfnamefont{J.}~\bibnamefont{Machta}},
  \emph{\bibinfo{title}{Yucesoy, {K}atzgraber, and {M}achta reply:}},
  \bibinfo{journal}{Phys. Rev. Lett.} \textbf{\bibinfo{volume}{110}},
  \bibinfo{pages}{219702} (\bibinfo{year}{2013}).

\bibitem[{\citenamefont{Wang et~al.}(2014)\citenamefont{Wang, Machta, and
  Katzgraber}}]{Wang:TBC}
\bibinfo{author}{\bibfnamefont{W.}~\bibnamefont{Wang}},
  \bibinfo{author}{\bibfnamefont{J.}~\bibnamefont{Machta}}, \bibnamefont{and}
  \bibinfo{author}{\bibfnamefont{H.~G.} \bibnamefont{Katzgraber}},
  \emph{\bibinfo{title}{Evidence against a mean-field description of
  short-range spin glasses revealed through thermal boundary conditions}},
  \bibinfo{journal}{Phys. Rev. B} \textbf{\bibinfo{volume}{90}},
  \bibinfo{pages}{184412} (\bibinfo{year}{2014}).

\bibitem[{\citenamefont{Wang et~al.}(2018)\citenamefont{Wang, Wallin, and
  Lidmar}}]{Wang:EA4D}
\bibinfo{author}{\bibfnamefont{W.}~\bibnamefont{Wang}},
  \bibinfo{author}{\bibfnamefont{M.}~\bibnamefont{Wallin}}, \bibnamefont{and}
  \bibinfo{author}{\bibfnamefont{J.}~\bibnamefont{Lidmar}},
  \emph{\bibinfo{title}{{Chaotic temperature and bond dependence of
  four-dimensional Gaussian spin glasses with partial thermal boundary
  conditions}}}, \bibinfo{journal}{Phys. Rev. E} \textbf{\bibinfo{volume}{98}},
  \bibinfo{pages}{062122} (\bibinfo{year}{2018}).

\bibitem[{\citenamefont{Hukushima and Iba}(2003)}]{Hukushima:PA}
\bibinfo{author}{\bibfnamefont{K.}~\bibnamefont{Hukushima}} \bibnamefont{and}
  \bibinfo{author}{\bibfnamefont{Y.}~\bibnamefont{Iba}}, in
  \emph{\bibinfo{booktitle}{{The Monte Carlo method in the physical sciences:
  celebrating the 50th anniversary of the Metropolis algorithm}}}, edited by
  \bibinfo{editor}{\bibfnamefont{J.~E.} \bibnamefont{Gubernatis}}
  (\bibinfo{publisher}{AIP}, \bibinfo{year}{2003}), vol. \bibinfo{volume}{690},
  p. \bibinfo{pages}{200}.

\bibitem[{\citenamefont{Zhou and Chen}(2010)}]{Zhou:PA}
\bibinfo{author}{\bibfnamefont{E.}~\bibnamefont{Zhou}} \bibnamefont{and}
  \bibinfo{author}{\bibfnamefont{X.}~\bibnamefont{Chen}}, in
  \emph{\bibinfo{booktitle}{{Proceedings of the 2010 Winter Simulation
  Conference (WSC)}}} (\bibinfo{publisher}{Springer},
  \bibinfo{address}{Baltimore MD}, \bibinfo{year}{2010}), p.
  \bibinfo{pages}{1211}.

\bibitem[{\citenamefont{Machta}(2010)}]{Machta:PA}
\bibinfo{author}{\bibfnamefont{J.}~\bibnamefont{Machta}},
  \emph{\bibinfo{title}{{Population annealing with weighted averages: A {M}onte
  {C}arlo method for rough free-energy landscapes}}}, \bibinfo{journal}{Phys.
  Rev. E} \textbf{\bibinfo{volume}{82}}, \bibinfo{pages}{026704}
  (\bibinfo{year}{2010}).

\bibitem[{\citenamefont{Wang et~al.}(2015{\natexlab{a}})\citenamefont{Wang,
  Machta, and Katzgraber}}]{Wang:PA}
\bibinfo{author}{\bibfnamefont{W.}~\bibnamefont{Wang}},
  \bibinfo{author}{\bibfnamefont{J.}~\bibnamefont{Machta}}, \bibnamefont{and}
  \bibinfo{author}{\bibfnamefont{H.~G.} \bibnamefont{Katzgraber}},
  \emph{\bibinfo{title}{{Population annealing: Theory and application in spin
  glasses}}}, \bibinfo{journal}{Phys. Rev. E} \textbf{\bibinfo{volume}{92}},
  \bibinfo{pages}{063307} (\bibinfo{year}{2015}{\natexlab{a}}).

\bibitem[{\citenamefont{Barash et~al.}(2017)\citenamefont{Barash, Weigel,
  Borovský, Janke, and Shchur}}]{Weigel:PA}
\bibinfo{author}{\bibfnamefont{L.~Y.} \bibnamefont{Barash}},
  \bibinfo{author}{\bibfnamefont{M.}~\bibnamefont{Weigel}},
  \bibinfo{author}{\bibfnamefont{M.}~\bibnamefont{Borovský}},
  \bibinfo{author}{\bibfnamefont{W.}~\bibnamefont{Janke}}, \bibnamefont{and}
  \bibinfo{author}{\bibfnamefont{L.~N.} \bibnamefont{Shchur}},
  \emph{\bibinfo{title}{{GPU accelerated population annealing algorithm}}},
  \bibinfo{journal}{Computer Physics Communications}
  \textbf{\bibinfo{volume}{220}}, \bibinfo{pages}{341} (\bibinfo{year}{2017}).

\bibitem[{\citenamefont{Amey and Machta}(2018)}]{Amey:PA}
\bibinfo{author}{\bibfnamefont{C.}~\bibnamefont{Amey}} \bibnamefont{and}
  \bibinfo{author}{\bibfnamefont{J.}~\bibnamefont{Machta}},
  \emph{\bibinfo{title}{Analysis and optimization of population annealing}},
  \bibinfo{journal}{Phys. Rev. E} \textbf{\bibinfo{volume}{97}},
  \bibinfo{pages}{033301} (\bibinfo{year}{2018}).

\bibitem[{\citenamefont{Barzegar et~al.}(2018)\citenamefont{Barzegar, Pattison,
  Wang, and Katzgraber}}]{Amin:PA}
\bibinfo{author}{\bibfnamefont{A.}~\bibnamefont{Barzegar}},
  \bibinfo{author}{\bibfnamefont{C.}~\bibnamefont{Pattison}},
  \bibinfo{author}{\bibfnamefont{W.}~\bibnamefont{Wang}}, \bibnamefont{and}
  \bibinfo{author}{\bibfnamefont{H.~G.} \bibnamefont{Katzgraber}},
  \emph{\bibinfo{title}{{Optimization of population annealing Monte Carlo for
  large-scale spin-glass simulations}}}, \bibinfo{journal}{Phys. Rev. E}
  \textbf{\bibinfo{volume}{98}}, \bibinfo{pages}{053308}
  (\bibinfo{year}{2018}).

\bibitem[{\citenamefont{Katzgraber et~al.}(2006)\citenamefont{Katzgraber,
  K\"orner, and Young}}]{katzgraber:06}
\bibinfo{author}{\bibfnamefont{H.~G.} \bibnamefont{Katzgraber}},
  \bibinfo{author}{\bibfnamefont{M.}~\bibnamefont{K\"orner}}, \bibnamefont{and}
  \bibinfo{author}{\bibfnamefont{A.~P.} \bibnamefont{Young}},
  \emph{\bibinfo{title}{{Universality in three-dimensional Ising spin glasses:
  A Monte Carlo study}}}, \bibinfo{journal}{Phys. Rev. B}
  \textbf{\bibinfo{volume}{73}}, \bibinfo{pages}{224432}
  (\bibinfo{year}{2006}).

\bibitem[{\citenamefont{Wang et~al.}(2015{\natexlab{b}})\citenamefont{Wang,
  Machta, and Katzgraber}}]{Wang:GS}
\bibinfo{author}{\bibfnamefont{W.}~\bibnamefont{Wang}},
  \bibinfo{author}{\bibfnamefont{J.}~\bibnamefont{Machta}}, \bibnamefont{and}
  \bibinfo{author}{\bibfnamefont{H.~G.} \bibnamefont{Katzgraber}},
  \emph{\bibinfo{title}{{Comparing Monte Carlo methods for finding ground
  states of Ising spin glasses: Population annealing, simulated annealing, and
  parallel tempering}}}, \bibinfo{journal}{Phys. Rev. E}
  \textbf{\bibinfo{volume}{92}}, \bibinfo{pages}{013303}
  (\bibinfo{year}{2015}{\natexlab{b}}).

\bibitem[{\citenamefont{Katzgraber et~al.}(2001)\citenamefont{Katzgraber,
  Palassini, and Young}}]{katzgraber:01}
\bibinfo{author}{\bibfnamefont{H.~G.} \bibnamefont{Katzgraber}},
  \bibinfo{author}{\bibfnamefont{M.}~\bibnamefont{Palassini}},
  \bibnamefont{and} \bibinfo{author}{\bibfnamefont{A.~P.} \bibnamefont{Young}},
  \emph{\bibinfo{title}{{M}onte {C}arlo simulations of spin glasses at low
  temperatures}}, \bibinfo{journal}{Phys. Rev. B}
  \textbf{\bibinfo{volume}{63}}, \bibinfo{pages}{184422}
  (\bibinfo{year}{2001}).

\bibitem[{\citenamefont{Marinari et~al.}(2000)\citenamefont{Marinari, Parisi,
  Ricci-Tersenghi, Ruiz-Lorenzo, and Zuliani}}]{Enzo:Review}
\bibinfo{author}{\bibfnamefont{E.}~\bibnamefont{Marinari}},
  \bibinfo{author}{\bibfnamefont{G.}~\bibnamefont{Parisi}},
  \bibinfo{author}{\bibfnamefont{F.}~\bibnamefont{Ricci-Tersenghi}},
  \bibinfo{author}{\bibfnamefont{J.~J.} \bibnamefont{Ruiz-Lorenzo}},
  \bibnamefont{and} \bibinfo{author}{\bibfnamefont{F.}~\bibnamefont{Zuliani}},
  \emph{\bibinfo{title}{{Replica Symmetry Breaking in Short-Range Spin Glasses:
  Theoretical Foundations and Numerical Evidences}}}, \bibinfo{journal}{Journal
  of Statistical Physics} \textbf{\bibinfo{volume}{98}}, \bibinfo{pages}{973}
  (\bibinfo{year}{2000}).

\bibitem[{\citenamefont{Palassini and Young}(2000)}]{palassini:00}
\bibinfo{author}{\bibfnamefont{M.}~\bibnamefont{Palassini}} \bibnamefont{and}
  \bibinfo{author}{\bibfnamefont{A.~P.} \bibnamefont{Young}},
  \emph{\bibinfo{title}{Nature of the spin glass state}},
  \bibinfo{journal}{Phys. Rev. Lett.} \textbf{\bibinfo{volume}{85}},
  \bibinfo{pages}{3017} (\bibinfo{year}{2000}).

\bibitem[{\citenamefont{{J. M. Hammersley and D. C. Handscomb}}(1964)}]{HH:MC}
\bibinfo{author}{\bibnamefont{{J. M. Hammersley and D. C. Handscomb}}},
  \emph{\bibinfo{title}{{Monte Carlo Methods}}}, {Monographs on Statistics and
  Applied Probability} (\bibinfo{publisher}{{Springer Netherlands}},
  \bibinfo{year}{1964}).

\bibitem[{\citenamefont{Liu}(2004)}]{Liu2004}
\bibinfo{author}{\bibfnamefont{J.~S.} \bibnamefont{Liu}},
  \emph{\bibinfo{title}{{Monte Carlo Strategies in Scientific Computing}}},
  Springer Series in Statistics (\bibinfo{publisher}{Springer New York},
  \bibinfo{address}{New York, NY}, \bibinfo{year}{2004}).

\bibitem[{\citenamefont{Fernandez and Martin-Mayor}(2009)}]{Fernandez:MC}
\bibinfo{author}{\bibfnamefont{L.~A.} \bibnamefont{Fernandez}}
  \bibnamefont{and}
  \bibinfo{author}{\bibfnamefont{V.}~\bibnamefont{Martin-Mayor}},
  \emph{\bibinfo{title}{{Mean-value identities as an opportunity for Monte
  Carlo error reduction}}}, \bibinfo{journal}{Phys. Rev. E}
  \textbf{\bibinfo{volume}{79}}, \bibinfo{pages}{051109}
  (\bibinfo{year}{2009}).

\bibitem[{\citenamefont{Boettcher}(2005)}]{boettcher:05d}
\bibinfo{author}{\bibfnamefont{S.}~\bibnamefont{Boettcher}},
  \emph{\bibinfo{title}{{Stiffness of the {E}dwards-{A}nderson Model in all
  Dimensions}}}, \bibinfo{journal}{Phys. Rev. Lett.}
  \textbf{\bibinfo{volume}{95}}, \bibinfo{pages}{197205}
  (\bibinfo{year}{2005}).

\bibitem[{\citenamefont{Aspelmeier et~al.}(2016)\citenamefont{Aspelmeier, Wang,
  Moore, and Katzgraber}}]{Wang:KAS}
\bibinfo{author}{\bibfnamefont{T.}~\bibnamefont{Aspelmeier}},
  \bibinfo{author}{\bibfnamefont{W.}~\bibnamefont{Wang}},
  \bibinfo{author}{\bibfnamefont{M.~A.} \bibnamefont{Moore}}, \bibnamefont{and}
  \bibinfo{author}{\bibfnamefont{H.~G.} \bibnamefont{Katzgraber}},
  \emph{\bibinfo{title}{{Interface free-energy exponent in the one-dimensional
  Ising spin glass with long-range interactions in both the droplet and broken
  replica symmetry regions}}}, \bibinfo{journal}{Phys. Rev. E}
  \textbf{\bibinfo{volume}{94}}, \bibinfo{pages}{022116}
  (\bibinfo{year}{2016}).

\end{thebibliography}

\end{document}